\documentclass[preprint2]{aastex}

\slugcomment{Based on observations made at San Pedro Martir Observatory.}

\shorttitle{Kinematics and dynamics in \objectname{NGC 4449}}
\shortauthors{Valdez--Guti\'errez et al.}

\begin{document}

\title{Unveiling the Kinematics and Dynamics of Ionized gas in
the Nearby Irregular Galaxy \objectname{NGC 4449}}

\author{
Margarita Valdez--Guti\'errez\altaffilmark{1,2,3},
Margarita Rosado\altaffilmark{4},
Iv\^anio Puerari\altaffilmark{1},
Leonid Georgiev\altaffilmark{4},
Jordanka Borissova\altaffilmark{5},
and
Patricia Ambrocio--Cruz\altaffilmark{4}
}

\altaffiltext{1}{
Instituto Nacional de Astrof\'\i sica, Optica y Electr\'onica; Calle Luis
Enrique Erro 1, 72840, Tonantzintla, Puebla, M\'exico; mago@inaoep.mx,
puerari@inaoep.mx
}

\altaffiltext{2}{
Current Address: Observatoire de Paris--LERMA, 61, Avenue de l'Observatoire, 
75014, Paris, France; margarita.valdez@obspm.fr 
}

\altaffiltext{3}{
Postdoctoral CONACyT Fellow
}

\altaffiltext{4}{
Instituto de Astronom\'\i a, UNAM; Apartado Postal 70--264, 04510,
M\'exico, D. F., M\'exico; margarit@astroscu.unam.mx,
georgiev@astroscu.unam.mx, patricia@astroscu.unam.mx
}

\altaffiltext{5}{
Institute of Astronomy, Bulgarian Academy of Sciences and Isaac Newton
Institute of Chile Bulgarian Branch, Bulgaria; 72 Tsarigradsko
chauss\`ee, BG --1784 Sofia, Bulgaria; jura@haemimont.bg
}

\begin{abstract}

A detailed kinematic analysis of ionized gas in the nearby irregular 
galaxy \objectname{NGC 4449} is presented. Observations are conducted 
in the spectral lines of H{$\alpha$} and [\ion{S}{2}].
Our scanning Fabry--Perot interferometric observations are presented 
from both a global as well as a local perspective.
We have analysed the global velocity field, the spatially extended 
diffuse gaseous component (DIG), the \ion{H}{2} region populations, 
and, furthermore, have determined the rotation curve based on the 
heliocentric radial velocities of the global H{$\alpha$} spatial 
distribution.

Our results for \objectname{NGC 4449} show that the optical velocity 
field has a decreasing value in radial velocity along the optical bar 
from NE to SW, presenting an anticorrelation relative to the outer 
velocity field of the \ion{H}{1} component. This is in agreement with 
previous studies.
The DIG component that permeates the entire galaxy was analysed (up
to a limiting surface brightness of $\sim 3.165\times10^{-5}$ ergs 
cm$^{-2}$ s$^{-1}$ steradian$^{-1}$) in terms of its radial velocity 
field as well as its velocity dispersions. 
We find that the diffuse gas component presents peculiar kinematical 
features such as abrupt velocity gradients and highly supersonic 
velocity dispersions ($\sigma\sim$ 4 times the values of the nearest 
\ion{H}{2} regions) but that its kinematical and dynamical influence 
is important on both global and local scales.
The optical rotation curve of this nearby irregular shows that the NE 
sector rotates like a solid body ($V_{rot}\sim$40 km s$^{-1}$ at $R=$2 kpc). 
In contrast, for the SW side, our results are not conclusive; the behavior 
of the gas at those locations is chaotic.
We conclude that the origin of such complex kinematics and dynamics is 
undoubtedly related to the aftermath of an interaction experienced by 
this galaxy in the past.

\end{abstract}


\section{Introduction}

In terms of frequency of occurrence, the most common morphology for
galaxies in the Universe are the Irregular (Irr) strain
\citep{gallagher84}. Irregulars may serve as morphological
Rosetta stones bridging the low and high redshift Universe
\citep{block01}. Historically, drawings of Irrs may be
found as early as 1847 (the Large Magellanic Cloud, classified by Hubble
as an irregular, was sketched by Sir John Herschel in that year).
When imaged in the near--infrared, a handful of irregulars betray a
magnificent -- albeit weak -- spiral density wave; examples are
NGC 5195 and NGC 922 (the latter galaxy even shows spiral arm
modulation). Generally, however, there is no decoupling of the
Population I and Population II disks; many optical irregulars remain
irregular in the near--infrared. One of the most intriguing 
questions yet to be answered is exactly what triggers/drives the 
formation of stars in galaxies.

Here the Irr species offer unprecedented opportunities, for
they are excellent laboratories in which one can examine the
star formation process --and its influence on the interstellar medium 
(ISM)-- often in the complete absence of spiral density waves.
Furthermore, the existence of fossil interacting features in
some of these galaxies makes them worthy of detailed study 
\citep{hunter00}.
The global study of the kinematical behavior of the gas in 
such environments is dynamically important. The presence of embedded 
superbubbles, giant \ion{H}{2} regions, filaments, supernova remnants 
(SNRs) and their combined effects, makes the global velocity field of
irregular galaxies both stimulating and important to analyze.

The face of the galaxy \objectname{NGC 4449}, a nearby Magellanic 
type Irr galaxy, is highly peculiar, with H${\alpha}$ streamers and 
filaments everywhere. Not only does \objectname{NGC 4449} present
giant \ion{H}{2} regions, but also a SNR, superbubbles,
and a diffuse emission (which embraces almost the entire optical 
boundary). The faint H${\alpha}$ filament population alluded to above 
is intriguing; the filaments delineate the inner edge of a
2 kpc diameter \ion{H}{1} supershell \citep{hunter97} located at 
the northwest side of the main optical body. At other wavelengths, 
there is a body of evidences suggesting a close connection between the 
diffuse nebulae, the diffuse emission and the X--ray emission from 
the center to the west of the main body
\citep{dellaceca97,vogler97,bomans97}.
As in the Large Magellanic Cloud, the galaxy \objectname{NGC 4449} 
presents a  bar--like morphology; it is located at a distance of 3.7 Mpc 
\citep{bajaja94}. \objectname{NGC4449} is not isolated in the sky; 
it is a member of the CVn I Cloud which includes a number of Im 
dwarfs and galaxies such as NGC 4214, NGC 4244 and IC 4182. For a list 
of relevant parameters, the reader is referred to Table~\ref{parameters}.

\objectname{NGC 4449} shows an unusually rich distribution of blue
supergiant stars at its northern periphery \citep{hunter97b};
this conclusion has been enhanced by ultraviolet (UV) imaging \citep{hill98}.
The UV results indicate that the UV--emitting populations (OB complexes) 
lie in the northern part of \objectname{NGC 4449}, while direct propagation 
of star formation appears possible south of the bar.
The latter conclusion was reached using the H$\alpha$/FUV(2231\AA) ratio,
from which age estimations of the OB complexes of this galaxy were derived, 
indicative of gradients in stellar age. The direct propagation of star 
formation in \objectname{NGC 4449} places this galaxy as one of a handful
--besides our \objectname{Galaxy}, \objectname{M82} \citep{satyapal97} and 
a couple of \objectname{LMC} OB associations and nebulae 
\citep{parker92,laval92,rosado96}-- where sequential star formation has been
discerned and measured \citep{hill98}.

The morphological and dynamical peculiarities in \objectname{NGC 4449}
are also reflected in its \ion{H}{1} and ionized gas contents.
The neutral gas in the outer parts of \objectname{NGC 4449} 
\citep{vanwoerden75, hunter98} seems to be counter rotating relative to the 
ionized gas distributed in the inner disk \citep{sabbadin84,valdez00}.
This counter rotation implies the co--existence of two physical systems
rotating in opposite directions in this galaxy. In Irrs, this behavior has
only be seen in a few examples: \objectname{IC 10} \citep{shostak89} and 
\objectname{NGC 4449} \citep{sabbadin84}.
As does the LMC, \objectname{NGC 4449} has a significant magnetic field, a 
fact which has only recently emerged \citep{otmianowska00,chyzy00}.
One of the conclusions reached in these studies is that the existence of a
global magnetic field is possible even for small values of galactic rotation 
(v$_{rot}$ $<$ 30 km s$^{-1}$).

The relatively small distance of \objectname{NGC 4449} earmarks this
specimen for detailed kinematical studies. The first Fabry--Perot (FP)
interferometric analysis of this galaxy was performed photographically
with a fixed gap FP etalon by \citet{crillon69}, who determined radial 
velocities for the more prominent complexes of \ion{H}{2} regions in
a spatial field of 5$\times$5 arcmin.
Other researchers who employed scanning FPs to study the kinematics
of the gas in \objectname{NGC 4449} are \citet{malumuth86}, 
\citet{arsenault88} and \citet{fuentes00}.
The observations in \citet{malumuth86} were spatially limited (3 arcmin 
square area) and kinematical interpretations were aggravated due to a loss 
of parallelism in the etalon plates.
\citet{arsenault88} and \citet{fuentes00} focused their work on the study of 
supersonic velocity dispersions in the brightest \ion{H}{2} regions
and those \ion{H}{2} regions belonging exclusively to the bar.
Finally, conventional kinematic studies by means of classic, long slit
spectroscopy for \objectname{NGC 4449} abound in the literature;
many projects have been devoted to the most prominent \ion{H}{2} regions as 
well as to the filament and superbubble populations \citep[and references 
therein]{sabbadin84,hartmann86,hunter97,martin98}.

In the present analysis, we revisit the analysis of the kinematics of the 
ionized gas in \objectname {NGC 4449} as a part of a long term project 
focused on the study of the interrelationship between gas and stars in 
irregular galaxies 
\citep[and references therein]{valdez01, rosado01, borissova00}.
This study presents, for the first time, a tridimensional kinematical 
analysis of the entire, optical distribution of the ionized gas in 
\objectname{NGC 4449}. 
Earlier studies had been undertaken by means of longslit spectroscopy and 
FP interferometry, but focused only on localised objects (such as ionised 
hydrogen regions) or otherwise, in a smaller field of view compared to the 
present work.

In Section 2 we describe the FP observations and data reduction. The
global, optical morphology in four optical lines are presented in Section 3, 
as is a statistical analysis of the \ion{H}{2} region population.
The kinematics of local (\ion{H}{2} regions) and global scale features 
(diffuse ionized gas, global velocity field) are discussed in Section 4. 
Section 5 is devoted to dynamical results -- as deduced from our optical
rotation curve. In Section 6 we compare our kinematical results with 
published \ion{H}{1} ones. Finally, in Section 7, we present our conclusions.

\section{Observations and Data Reduction}

The observations were carried out during four runs with the UNAM Scanning
Fabry--Perot interferometer PUMA, attached to the f/7.9 Ritchey--Chr\'etien
focus of the 2.1 m\ telescope at the Observatorio Astron\'omico Nacional at 
San Pedro M\'artir, B.C., M\'exico. The log of the observations is described 
in Table~\ref{observations}.

The PUMA setup is composed of a scanning FP interferometer, a focal
reducer with a f/3.95 camera, a filter wheel, a calibration system and a
Tektronix 1K$\times$1K CCD detector \citep{rosado95}. Other components of the 
instrumental setup used in each run, are the same as described in Table 2 of 
\citet{valdez01}. Depending on the CCD reading (bin = 1, 2 or 4), the pixel 
size is 0\farcs58, 1\farcs 17 or 2\farcs34, respectively. 
At our assumed distance (see Table~\ref{parameters}), the pixel linear--size 
is equivalent to 10.5, 21 and 42 pc, respectively. A 10$\arcmin$ field of view 
is spanned, embracing the entire optical extent \objectname{NGC 4449} out to 
the de Vaucouleurs 25 mag arcsec$^{-2}$ isophote.

We obtained a set of direct images --in three runs-- in the lines of H$\alpha$,
[\ion{S}{2}] ($\lambda$ 6717, 6731 \AA), [\ion{N}{2}] ($\lambda$ 6584 \AA)
and [\ion{O}{3}] ($\lambda$ 5007 \AA) using the PUMA in its direct image
mode, i.e. without the FP on the optical axis. For their calibration, we have
observed, under the same conditions, the spectrophotometric standard star
BD+25$\arcdeg$3941 \citep{barnes82}. 
In the fourth run we exclusively used the H$\alpha$ filter for direct imaging: 
see Section 3 and Table~\ref{observations} for more details of these 
observations.
In the third run, the seeing in the direct images of \objectname{NGC 4449}
--images which were later used to perform our flux calibration-- was 
of the order of 2$\arcsec$ (FWHM).

In addition, we have obtained scanning FP data cubes in the lines of H$\alpha$
and [\ion{S}{2}]. The FP data cubes are composed of 48 steps covering the 
entire spectral range, with integration times of 60 seconds per step. In this 
configuration, the final data cubes had dimensions of 512$\times$512$\times$48 
(after applying a binning of two).
We have obtained 2 data cubes per filter; the total exposure time was 1.6 
hours in each case. The measured seeing in these data cubes is of the order of
2\farcs7 (FWHM).

FP data reductions (phase calibration and kinematical analysis) were carried
out in the same manner as in \citet{valdez01}, mainly using the software
package CIGALE \citep{lecoarer93}. The photometry was performed under
IRAF\footnote{IRAF is distributed by the National Optical Astronomy
Observatories, operated by the Association of
Universities for Research in Astronomy, Inc., under
cooperative agreement with the National Science
Foundation.}.
Our data cubes in H$\alpha$ are contaminated with night--sky lines such as
the lines of OH at 6568.78 \AA\ and 6568.77 \AA.
These night--sky lines fall inside the main emission of \objectname{NGC 4449},
centered at channel 14 or at the heliocentric velocity +295 km s$^{-1}$ (in 
the redshifted wing of single velocity profiles). Such night--sky lines have 
been subtracted using an interactive routine of CIGALE and consequently
our radial velocity dispersions are free of them. The [\ion{S}{2}] data cubes 
are less contaminated by night--sky lines; however, the S/N ratio is not as 
good as for the H$\alpha$ cubes and are only used to compare and to complement 
the H$\alpha$ observations.

To extract the kinematic information from the FP data cubes, we adopt the same 
methodology described in \citet{valdez01}. The radial velocity profiles were 
fitted by Gaussian functions once deconvolved with the instrumental function 
(an Airy function). We also constructed a set of maps, or moment maps.
The first moment is given by the radial velocity map, which
CIGALE calculates by finding the barycenter in the radial velocity profile of 
each pixel. Cleaning this map at 2 $\sigma$ times  the value of the standard
background emission, ensures that the remaining features are galactic.
The second moment map or the velocity dispersion map, was derived using the
Adhoc package \citep{boulesteix93}.
Besides our moment maps, CIGALE obtained a monochromatic map after integration
of the radial velocity profile (velocity versus intensity along the 48 
channels) of each pixel, up to a certain fraction of the peak (usually 70\%).
This map enables us to separate the monochromatic emission from the continuum
\citep{lecoarer93}.

\subsection{Photometric Calibration}

To accomplish the flux calibration of our H$\alpha$ data, the direct images
of the standard star BD+25$\arcdeg$3941 and the direct images of the galaxy
(see previous Section and Table~\ref{observations} for details) were
reduced and analysed in the usual way.
In this manner, our conversion factor (to secure  absolute calibrated
surface brightnesses) was:
1 count s$^{-1}$ pix$^{-1}$ = 1.266$\times$10$^{-4}$ erg cm$^{-2}$ s$^{-1}$
steradian$^{-1}$, which in units of flux is: 1.018$\times$10$^{-15}$ erg
cm$^{-2}$ s$^{-1}$. 
For images binned by a factor of two,  1 count s$^{-1}$ pix$^{-1}$ 
corresponds to 3.165$\times$10$^{-5}$ erg cm$^{-2}$ s$^{-1}$ steradian$^{-1}$.

Given that \objectname{NGC 4449} presents a conspicuous diffuse component
pervading almost all of its disk and taking into account that this
component is more intense towards the NE--SW bar \citep{scowen92}, we
considered it appropriate to subtract such diffuse emission only
on our direct images, in order to determine the \ion{H}{2} region diameters 
more accurately.
For the datacubes, we used another scheme to separate the diffuse
emission; that methodology will be explained in the next section.

\begin{figure*}[htb]
\includegraphics{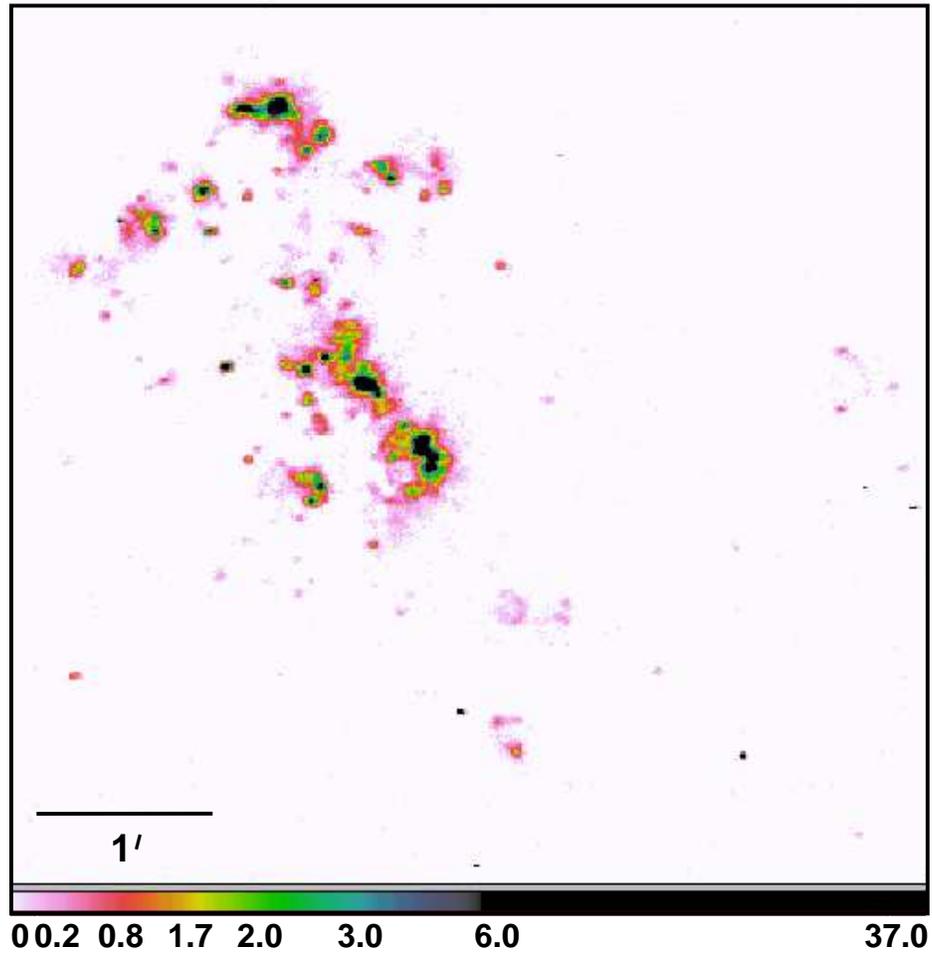}
\vspace{10.6cm}
\caption{A flux--calibrated H$\alpha$ image of NGC 4449, 
free of diffuse emission (see text for further details). The colors are 
related to counts sec$^{-1}$.
In order to secure H$\alpha$ fluxes, luminosities and surface brightnesses,
multiplications by  factors of 1.018$\times$10$^{-15}$ erg cm$^{-2}$ s$^{-1}$,
1.66$\times$10$^{36}$ erg s$^{-1}$ and 1.266$\times$10$^{-4}$ erg cm$^{-2}$
s$^{-1}$ steradian$^{-1}$, respectively, are necessitated.
The scale is indicated at bottom left. North is up; East to the left.}
\label{cali_free_dig}
\end{figure*}

To subtract the diffuse emission in our H$\alpha$ direct image of
\objectname{NGC 4449}, we used the technique (well known in photometry) of 
median filtering.
In our case, the best filter was that with a square box of 97 pixels. This
filter has to be at least 3 times greater (in spatial area) than the
most extended feature to be analysed, in order to ensure reliable
H$\alpha$ counts (after subtraction of the filtered image from the direct one).
Once the diffuse emission was subtracted, the background was removed at 3
$\sigma$ of its mean value, to ensure galactic membership of the
remaining features. The resulting image is shown in Figure~\ref{cali_free_dig}.
However, some local background persisted, especially at the bar, where
the crowding is high. This effect can be seen in our 
Figure~\ref{cali_free_dig} and Figure 7 in \citet{fuentes00}.

\begin{figure*}[htb]
\includegraphics{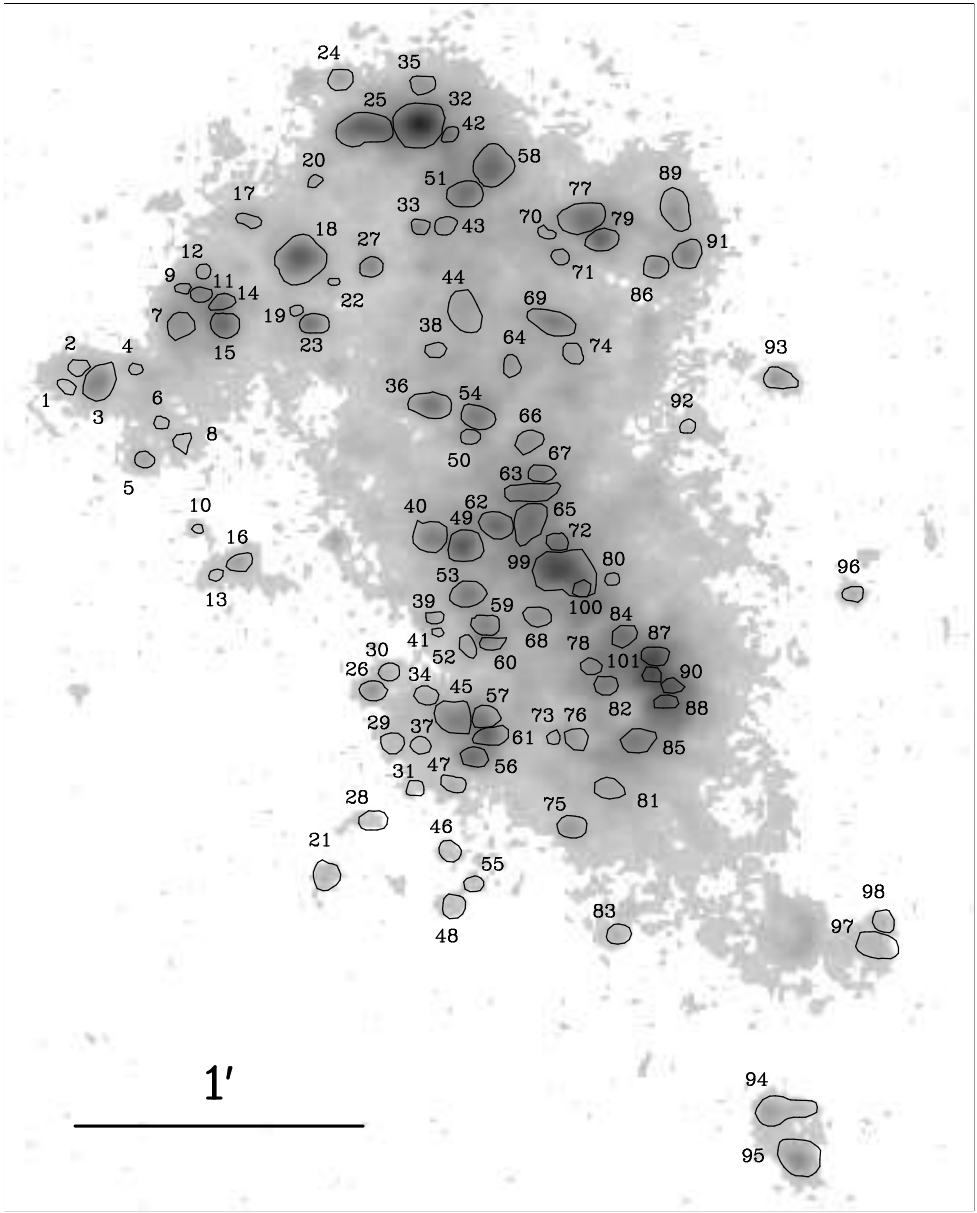}
\vspace{12.7cm}
\caption{Polygonal apertures (defined without diffuse emission; see
Figure~\ref{cali_free_dig}) were used to obtain H$\alpha$ fluxes 
from the \ion{H}{2} region population in NGC 4449.
In order to display weak features, we adopted a logarithmic scale for 
the gray levels. See text for further details. Numbers 
correspond to our ``M" sources in Table~\ref{parameters_1}. 
The scale is indicated at bottom left. North is up and East to the
left.}
\label{apertures}
\end{figure*}

The next step was to define the \ion{H}{2} region boundaries. Later,
these were used as apertures from which our photometry was performed. 
Polygonal apertures were chosen so as to contain all emission from a given
\ion{H}{2} region (Figure~\ref{apertures}).
Special care was taken with objects embedded in crowded regions, given that 
the limits between individual \ion{H}{2} regions and the local background 
were problematic, even when the diffuse emission had been eliminated.
In a crowded region (such as the bar) the definition of the apertures
was carried out by locally setting a contour map of the complex, to determine 
the relative contribution of each embedded \ion{H}{2} region.
Pixels within the apertures were then summed to generate the total area
and the total counts for every \ion{H}{2} region. The total H$\alpha$ flux for
each \ion{H}{2} region was obtained by multiplying the resulting counts
by the calibration conversion factor.

Our resulting H$\alpha$ fluxes for the \ion{H}{2} regions are presented
in column 4 in Table \ref{parameters_1}.
The uncertainty in the derived fluxes of crowded regions ranges from $\pm$10\%
for the brightest objects, to 30\% for the faintest \ion{H}{2} regions.
The lowest and highest levels for which we have reliable data in our cataloged
\ion{H}{2} regions are: 1.9$\times$10$^{37}$ and 3.38$\times$10$^{39}$ erg 
s$^{-1}$, respectively; these values correspond to a surface brightness of
6.309$\times$10$^{-5}$  and 1.047$\times$10$^{-3}$ erg cm$^{-2}$ s$^{-1}$
steradian$^{-1}$.
The lowest value corresponds to our region M92, and the highest one is found in
our region M32. The H$\alpha$ luminosity of the latter region (M32) is
approximately 1/5 that of the H$\alpha$ luminosity for 30 Doradus 
(1.5$\times$10$^{40}$ erg s$^{-1}$) and two orders of magnitude brighter than 
the Orion Nebula --Messier 42, 10$^{37}$ erg s$^{-1}$, see \citet{kennicutt84}.

At first glance, it may appear that removal of diffuse emission
would help us to assign definite diameters to areas of interest (e.g., 
ionised hydrogen regions) and hence compute fluxes coming from each such 
region of interest, but there are complexities, as discussed below.

Our H$\alpha$ fluxes free of diffuse emission were compared against that
obtained by \citet{fuentes00}, who also removed the diffuse emission in 
their work. Fuentes-Masip values resulted in brighter ones than ours, by 
an amount of $\sim$60\% in the most extreme case, in spite of their diameters 
being smaller than ours. Also, our best fit to Fuentes-Masip falls on a 
regression line parallel to the diagonal (slope$\sim0.7$).
Hence, in order to explore the importance as to whether we should subtract
the diffuse emission in the \ion{H}{2} region photometry, we compared the
fluxes obtained by \citet{hill98}(diffuse emission included) against those 
of Fuentes-Masip (free of diffuse emission).
The chosen \ion{H}{2} regions were SB89, 108, 114, 133, and 208. The difference
in log (L) was 0.6 in the worst case, with the Fuentes-Masip values
always brighter than Hill's; recall that Hill's apertures were larger, and had 
diffuse emission included.
In the light of such results, we decided it appropriate to apply our
polygonal apertures (defined without diffuse emission) on the monochromatic 
image (diffuse emission included and transformed to bin=1) and to compute 
fluxes.

\begin{figure}[ht]
\includegraphics{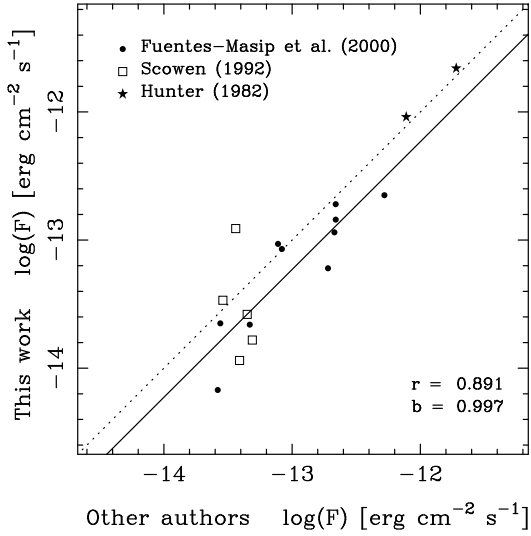}
\vspace{6.8cm}
\caption{A comparison between our flux determinations with  
published fluxes as determined in other investigations.
The solid line is our best fit to \citet{fuentes00}, while the dotted 
one represents the diagonal. Values of the correlation coefficient (r) 
and the slope (b) of the least squares fit are given at bottom right.}
\label{comparison}
\end{figure}

Our flux determinations were compared against published data by
\citet{hunter82}, \citet{scowen92} and \citet{fuentes00}. The regions chosen
to compare were selected from \citet{sabbadin79} and \citet{crillon69}
lists.
They also were chosen to be relatively isolated, round and intense (especially
in the case when a region was embedded in the bar).
Figure \ref{comparison} shows the comparison. In this figure
it can be seen that our fluxes are in excellent agreement with those of 
\citet{hunter82}, for two bright \ion{H}{2} complexes at the bar.
Our discrepancies with \citet{scowen92} probably are given by the uncertainty 
in the positions of the apertures.
With respect to \citet{fuentes00}, our comparison to their values results in a
fit with a slope value (b=0.99) parallel to the diagonal, where the
Fuentes-Masip values are in general, brighter than ours.
The correlation ($r$$\sim$0.9) found in such fit, is high. This fact
points towards the homogeneity in our data and a systematic effect produced by
the Fuentes-Masip method of subtracting the diffuse emission and the influence
of such effect in the acquisition of their fluxes.

To obtain luminosities, our fluxes were corrected for galactic foreground
extinction, even though in the direction of \objectname{NGC 4449}, it is 
negligible: A$_B$=0.2 mag \citep{devauco76,burstein78}. The effects of 
internal extinction were not taken into account; they are not crucial in 
kinematical studies \citep{hippelein86}.

\subsection{Definition of the Apertures in the FP Data}

The \ion{H}{2} region apertures --rectangular boxes in this case-- in which 
the kinematical analysis was performed, were assigned using different criteria 
compared to the scheme followed in the direct imaging mode. The reason is 
due to the fact that we did not want to apply any filtering technique on our 
datacubes to eliminate the diffuse emission. This could otherwise introduce
a systematic effect over the continuum level at each pixel, as a consequence 
of such filtering.
Therefore, as a first step --as is customary in the analysis of FP data-- 
the boundaries were determined checking all the velocity maps in which 
emission was detected for a given \ion{H}{2} region. In this way, by 
inspecting each 0.41 \AA\, we were sure that contamination by diffuse 
emission was not important. Moreover, we only took as `valid' detections, 
only those detected in two successive velocity maps.

In order to check if  our FP box boundaries were effectively free (and not 
contaminated) by diffuse emission, we superimposed our FP boxes on the
FP monochromatic image in which the \ion{H}{2} regions were clearly delineated 
(such monochromatic images were cleaned of diffuse emission, as previously 
described in subsection 2.1). We next superimposed our FP boxes on the 
filtered direct image.
The comparison was highly satisfactory; only in some cases did we need to 
marginally redefine our boxes.
For some \ion{H}{2} regions embedded in the bar, we secured local contours to 
examine the relative contribution of the different components and in this way, 
to assign our box boundaries. The boxes are shown in Section 4.2.

The resulting catalog of kinematic properties (Table~\ref{parameters_2})
differs from that in Table~\ref{parameters_1}; this may be ascribed to
different binnings in each case (see Table~\ref{observations} for details).

\section{Global Morphology of the Ionized Gas in NGC 4449}

\begin{figure*}[ht]
\includegraphics{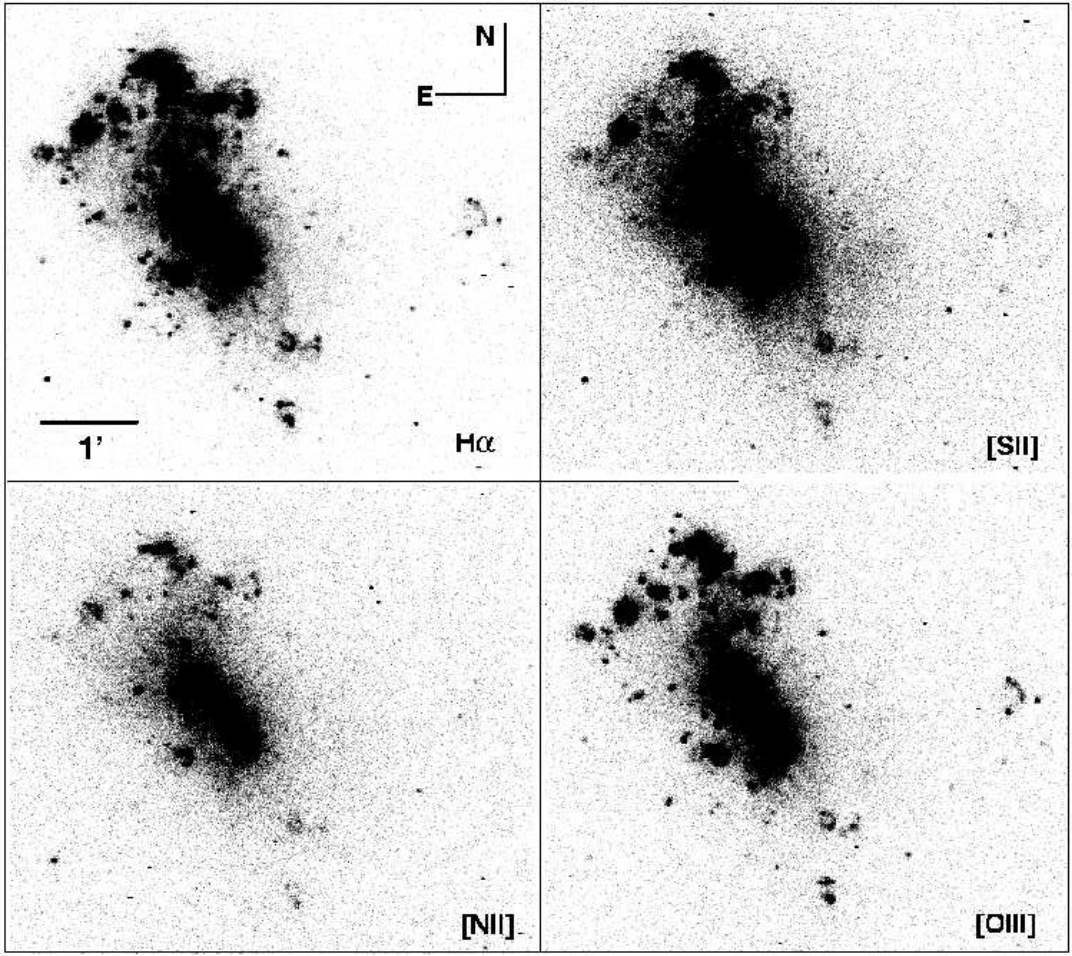}
\vspace{10.8cm}
\caption{ Direct images of NGC 4449 in
the lines of H$\alpha$, [\ion{S}{2}], [\ion{N}{2}]
and [\ion{O}{3}]. The exposure times are those reported in
Table~\ref{observations}. The scale and the orientation are indicated in the
first panel. }
\label{direct}
\end{figure*}

Figure \ref{direct} shows the emission of \objectname{NGC 4449} in the lines
of H$\alpha$, [\ion{S}{2}], [\ion{N}{2}] and [\ion{O}{3}] as obtained from
our direct images in the second run (see Section 2 and Table~\ref{observations}
for more details).
In these images it can be seen that the dimensions of the galaxy's main body
--at the adopted distance-- are approximately 4$\farcm$4 or 4.8 kpc and 
2$\farcm$7 or 3 kpc for its major and minor axes (not shown), respectively.
Seen at these large scales, the H$\alpha$ and [\ion{O}{3}] global
emission of \objectname{NGC 4449} appears to be mainly concentrated toward the
south, where copious numbers of \ion{H}{2} region complexes, superbubbles and
filaments are found.
However, the presence of filaments and superbubbles is not related
exclusively to the main body: an ``arc" of emission is visible at the west from
the photometric center as far as 2$\farcm$6 or 2.8 kpc. Such a feature is 
referred to in other works as arc 156--157 \citep{scowen92} or SGS3 
\citep{bomans97}, and it is better seen in H$\alpha$ and [\ion{S}{2}].

The most prominent spatial extent of emission \objectname{NGC 4449} is
located, in projection, along the bright bar -- which extends from NE to SW and
whose morphology is conspicuous in all the filters.
Along the bar lies a hugely strong continuum whose location
corresponds to the photometric center. This feature has been referred to
as a ``nucleus'' and its nature has been recently analysed by
\citet{boker01} and \citet{gelatt01}. The results of these works points 
towards the existence of a young super star--cluster embedded at that 
location: a fact which is possibly related to the previous interactions 
suffered by this galaxy.
Moreover, the bar possesses a pervasive diffuse ionized gas component.
This diffuse gas also embraces almost the entire optical extent of the
galaxy, as we have already alluded to.

To the north of \objectname{NGC 4449}, the bright complexes CM46 and
CM39 \citep{crillon69} emit more strongly in H$\alpha$, [\ion{S}{2}] and
[\ion{O}{3}] compared to [\ion{N}{2}], where they look dimmer.
The SNR previously reported \citep[also known as \ion{H}{2} region
SB187]{seaquist78,balick78} is located in our images at the SW border of our 
region M77 or V61 (see Tables \ref{parameters_1} and \ref{parameters_2}) and 
appears brighter in [\ion{O}{3}] than in H$\alpha$.
In our region M77(V61) the \ion{H}{2} regions SB185 and SB187 
\citep{sabbadin79} look almost blended and embedded in a diffuse emitting 
region. At the borders of this \ion{H}{2} complex, the presence of the 
superbubble SB2 \citep {bomans97} or \objectname{NGC 4449}--D 
\citep{martin98} is unmistakable as suggested by the complexity of the radial 
velocity field at that locale.
Apart from the features above, the galaxy appears to have a fuzzy morphology 
in the lines of [\ion{S}{2}] and [\ion{N}{2}], given that the continuum has 
not been subtracted. Nevertheless, the bar remains as the most distinct 
feature in contrast to other regions with negligible emission.

As far as the beautiful and sprawling filaments are concerned, these emerge 
in the H$\alpha$ and [\ion{O}{3}] filter images at first glance, as opposed 
through the [\ion{S}{2}] and [\ion{N}{2}] filters, where they are not well 
discerned (mainly due to the diffuse emission).
Filament 6 \citep{hunter97} and the bulk of filaments already reported 
\citep[and references therein]{bomans97}, are indeed prominent in our 
H$\alpha$ direct image. 
In the [\ion{O}{3}] image, the number of resolved filaments \citep{bomans97} 
is less, but filament 6 remains resolved.

We also have conducted H$\alpha$ direct observations around the locations where
some \ion{H}{1} maxima have been reported \citep{hunter98}. Information 
pertaining to this run is displayed in Tables~\ref{observations} and 
\ref{HI_maxima}.
In Table~\ref{observations}, the corresponding frames are indicated with 
positions 1, 2 and 3. Our goal was to look for traces of star formation at 
those locales. 
However, we did not find obvious H$\alpha$ emission at these three positions,
although we performed an unsharp masking treatment in our images. This result 
agrees with the work of \citet{hunter00b} for positions 2 and 3. At our 
position 1, there are no other published comparisons pertaining to the issue 
of star formation there.

\subsection{The \ion{H}{2} Region Population}

The \ion{H}{2} regions of \objectname{NGC 4449} have been catalogued by
\citet{crillon69}, \citet{hodge69}, \citet{sabbadin79} and \citet{hodge83b}.
The most extensive catalogue corresponds to \citet{sabbadin79} with 252 
regions. 
Our identifications are mainly based on this work and on the earlier work of 
\citet{hodge69}.

Our observations contain 140 \ion{H}{2} regions according to
\citet{sabbadin79}. Two other regions not catalogued in this work are
denoted in the second column of Table~\ref{parameters_1} and 
Table~\ref{parameters_2}, and demarcated as h61 and h66 according to 
\citet{hodge69}. Futhermore, regions denoted by a ``V" letter followed
by a number in column 2 of Table~\ref{parameters_1}, correspond to our sources 
listed in Table~\ref{parameters_2}.
The ionised hydrogen regions detected by \citet{sabbadin79} but not by us is 
due to several possibilities: very weak emission (and consequently not 
detected by us); alternatively, some were possibly not resolved with our 
telescopic and FP configuration.
Another issue is blending: in some cases, the boundaries between diffuse
\ion{H}{2} regions being part of one large complex were not obvious.
In such cases, we assigned them with a common boundary. Taking into account 
all these scenarios, only 84 and 101 regions were cleanly defined in our 
monochromatic and direct images respectively.

Table \ref{parameters_1} shows the results of our photometry performed on
the \ion{H}{2} regions in \objectname{NGC 4449}. Columns 1 and 2
present the identifications of each source in this work and in previous ones
\citep[and this work]{hodge69, sabbadin79}. Column 3 shows the equivalent
diameter (D), which is defined  in terms of the area (A) where the photometry 
is performed: A= $\pi$(D/2)$^2$. Column 4 displays the logarithm of the 
H$\alpha$ flux, derived as described in Section 2.1. Finally, columns 5 and 6 
show the luminosities and surface brightnesses.

\subsection{The \ion{H}{2} Region Luminosity Function}

\begin{figure}[htb]
\includegraphics{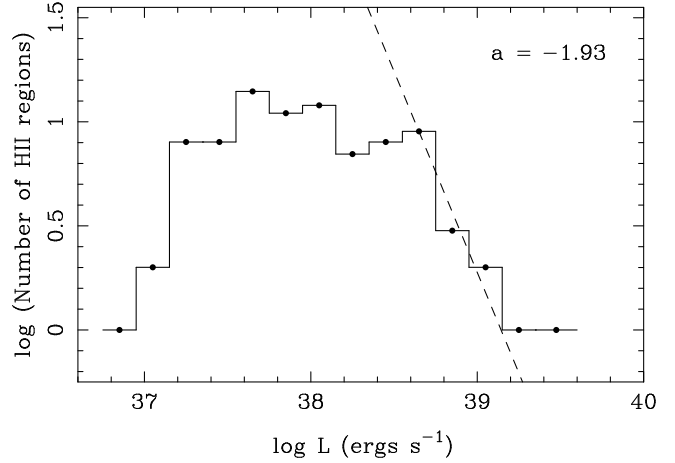}
\vspace{6.8cm}
\caption{ The H$\alpha$ luminosity function for the \ion{H}{2} regions in NGC 4449.
The dashed line is the least squares fit to a power law. The value of
the slope is given at top right.}
\label{lum_function}
\end{figure}

Figure \ref{lum_function} shows the H$\alpha$ luminosity function for the
\ion{H}{2} regions in \objectname{NGC 4449}. For several galaxies, it has been 
shown that the brighter luminosity end of the \ion{H}{2} region luminosity 
function can be fitted with a power law in the form: 
\mbox{$N(L)$=A$L^{a}$ $dL$ }, where $N$ is the number of \ion{H}{2} regions 
with luminosity $L$ in an interval of luminosity $dL$, and the slope $a$ is 
related to the morphological type of the galaxy \citep{kennicutt89}, ranging 
from mean values of --2.3 (Sab--Sb galaxies) to --1.75 (Irrs). It also has 
been suggested that different ranges of the luminosity function are related to
different kind of objects such as \ion{H}{2} regions ionized by single stars 
compared to those ionized by clusters of stars \citep{hodge89,youngblood99}.

In order to obtain the \ion{H}{2} region luminosity function, we followed the
procedure already described in \citet{valdez01}. For \objectname{NGC 4449} the 
bin width was more refined than in
\citet[with 30 \ion{H}{2} regions]{valdez01}, and the fit was calculated in 
bins brighter than $\log L= 38.6$, showing a negative slope.Composite regions 
as M87 or M90, were excluded in the fit. Our best fit slope for 
\objectname{NGC 4449} gives $a=-1.93\pm 0.2$. This value is in good agreement 
(within our error bars) with the value ($a=-1.9$) found by \citet{fuentes00}, 
and lie within the errors of the mean values found for a sample of Irr 
galaxies studied by \citet{kennicutt89}, \citet{strobel91}, 
\citet{kingsburgh98} and \citet{youngblood99}. These studies 
were performed on six, three, twenty--nine and four Irr galaxies respectively;
only the work by Kennicutt et al. includes \objectname{NGC 4449}.

\subsection{The \ion{H}{2} Region Diameter Distribution}

The \ion{H}{2} region diameter distribution can be expressed as:
\hbox{$N(>D)$=$N_0$\ exp${(-D/D_0)}$}, where \hbox{$N(>D$)} is the
cumulative function of \ion{H}{2} regions with diameter larger than $D$, 
and $D_0$ is the characteristic diameter.
This functional dependence was proposed by \citet{vandenbergh81} as an
universal law for all galaxy types and it has been used to study the 
statistics of the \ion{H}{2} region populations in combination with the 
\ion{H}{2} region luminosity function.
However, in a recent work \citep{pleuss00}, there appears to be a
dependence of the slope in the distribution function with respect to the 
spatial resolution. Thus, care has to be taken with regard to values in 
slope whose resolution data lies under 40 pc FWHM \citep{pleuss00}.

To calculate the distribution we have used the diameters of the \ion{H}{2}
regions measured on the direct H$\alpha$ images; these are to be found
in column 3 in Table \ref{parameters_1}.
According to our definition, the equivalent diameters range from 38.8 pc
(corresponding to our region M41) to 184.4 pc (for region M32), with a mean
value of 89.1$\pm$31.6 pc. Our mean diameter value is in agreement (within 
dispersion values and within adopted distance uncertainties) with a mean 
value previously reported \citep{fuentes00}. 
Recall that the latter study considers only 43 regions and that it also 
takes into account the subtraction of the diffuse emission in the 
determination of diameters. On the other hand, the large dispersion found in 
our mean equivalent diameter accounts for the wide range in the sizes of the
\ion{H}{2} regions in this galaxy.

\begin{figure}[ht]
\includegraphics{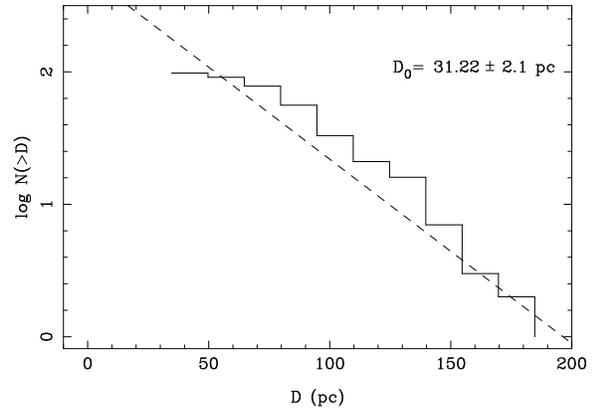}
\vspace{6.8cm}
\caption{The cumulative size distribution for the \ion{H}{2} regions in
NGC 4449, where the dashed straight line represents a least squares
fit. The value for the characteristic diameter is given in the top right 
hand corner.}
\label{distribution_HII}
\end{figure}

The resulting cumulative \ion{H}{2} region diameter distribution $N(>D)$ is
shown in Figure \ref{distribution_HII}, where we plot ${\rm log}\, N(>D)\,
vs\, D$. For \objectname{NGC 4449} a least squares fit gives 
$D_0=$1$\farcs$73$\pm$0.1 or 31.2$\pm$2.1 pc. In the calculation we have 
included the diameters of \ion{H}{2} complexes --not ring shaped-- which 
envelop more than one region. This value is in agreement with a previously 
published value by \citet{fuentes00} who obtained $D_0$=1$\farcs$6. 
However, our value differs from that of \citet {hodge83} and 
\citet{sabbadin79} who quoted larger values:
$D_0$= 5$\farcs$5 and 6$\farcs$2, respectively.
The disagreement here could possibly be due to the diffuse emission not 
being subtracted in these studies.
To this end, \citet{pleuss00} suggest that the intrinsic clustering properties
of the \ion{H}{2} regions plus diffuse emission and resolution effects, do 
indeed hamper reliable measurements of internal quantities (such as 
luminosities and diameters).

Pertaining to the correlation between $D_0$ and the absolute
magnitude $M_B$ given by \citet{strobel91}:

\mbox{${\rm log}D_0=0.25(\pm 0.29)-0.098(\pm 0.017) M_B$},

\noindent when we plot 
our observational value (${\rm log} D_0=1.5$) as well as the resulting 
value from the above equation, ${\rm log} D_0=2.06\pm0.3$, taking $M_B=-18.5$
\citep{schmidt92}, the first value does not fit the correlation in Figure 11
in \citet{strobel91}, while the latter does.
This supports the conclusion of \citet{strobel91} and \citet{youngblood99}, 
who analyzed a sample of 29 Irrs (\objectname{NGC 4449} not included) and 
Blue Compact Dwarfs, and who highlighted  the weakness of this correlation.

\section{Kinematical Analysis on Global and Local Scales}

\subsection{The Global Velocity Field of NGC 4449}

\begin{figure*}[htb]
\includegraphics{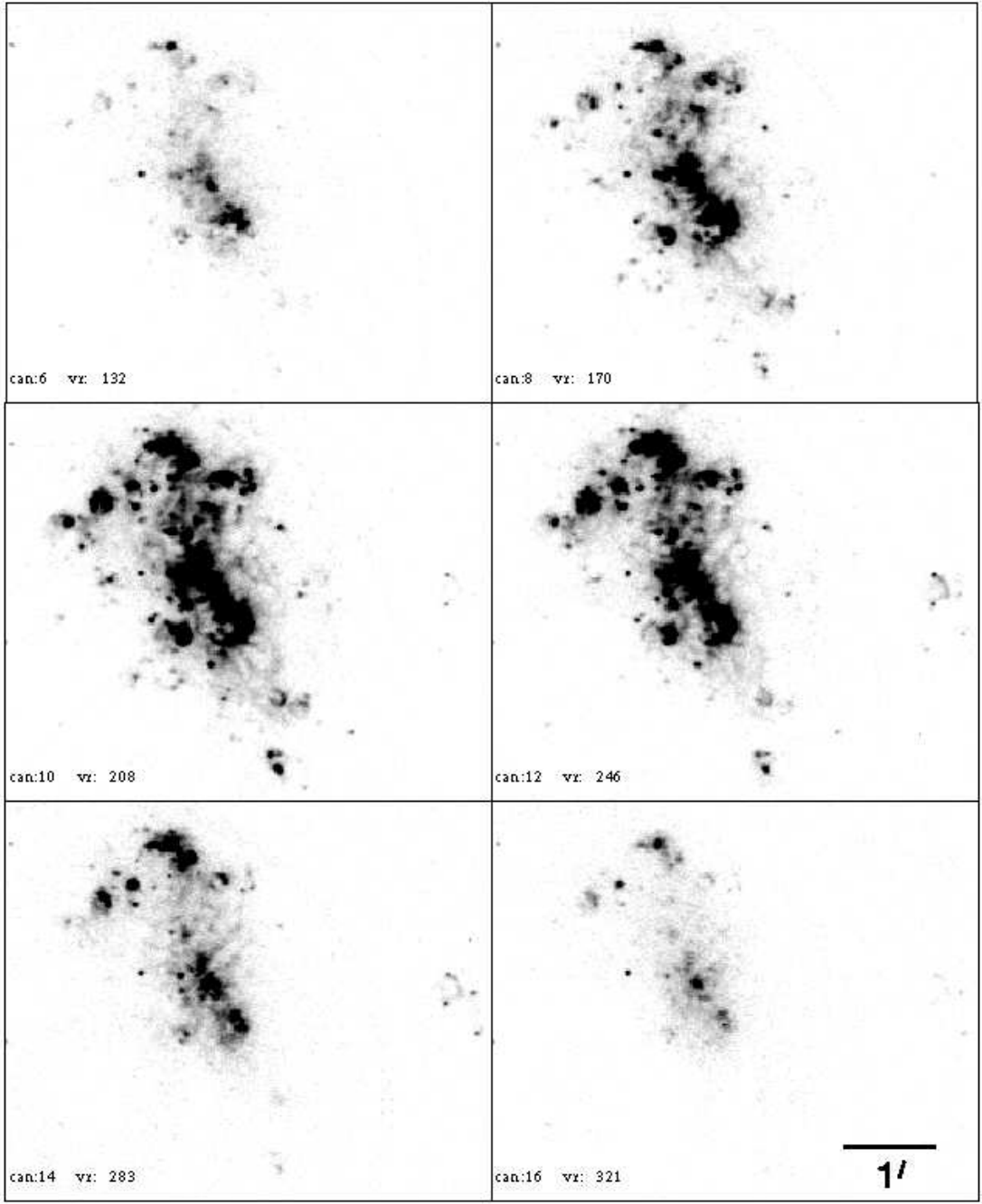}
\vspace{19.0cm}
\caption{NGC 4449 H$\alpha$ radial velocity maps. The channel number as
well as the heliocentric radial velocity (in km s$^{-1}$) are indicated in each
panel. The scale is indicated in the panel corresponding to channel 16, at
bottom right.}
\label{mosaic_vr}
\end{figure*}

\begin{figure*}[htb]
\includegraphics{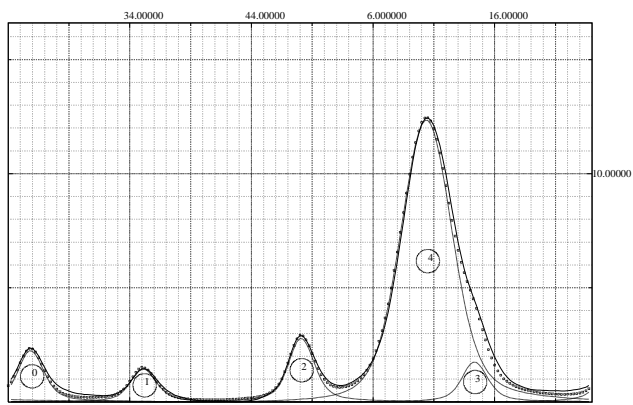}
\vspace{5.5cm}
\caption{The observed  H$\alpha$  global radial velocity distribution
in NGC 4449 (see text for details). The horizontal and vertical axes 
correspond to the channel number and intensity (arbitrary units), 
respectively. 
The fit to the sky--emission lines has been performed by means of
curve numbers 0, 1, 2 and 3 (see text for details).}
\label{vr_global}
\end{figure*}

In \objectname{NGC 4449} we detected H$\alpha$ emission ranging from 75 to
472 km s$^{-1}$ in heliocentric radial velocities; some velocity channels are
displayed in Figure~\ref{mosaic_vr}. Also, the global radial velocity 
profiles for all emission emanating from this galaxy has been fitted. 
The results are displayed in Figure~\ref{vr_global}. In this figure one can 
immediately see that the peak velocity of the whole emission is at channel 
10.5 (or 218 $\pm$ 4 km s$^{-1}$) with a corrected velocity dispersion of 
1.7 channels (or 31.5 $\pm$ 10 km s$^{-1}$).
The other curves in that Figure correspond to sky--line fittings; curve 3
contributes to the profile in our line under study (H$\alpha$), and it has
been taken into account in all calculations determining radial velocities and 
velocity dispersions.
The peak velocity can be considered the systemic velocity of the galaxy, and
the velocity dispersion can be visualized as the representative value of the
velocity dispersion of all the objects embedded in the disk.
At this point, we do however caution that H$\alpha$ velocity dispersions
are subject to many factors which contribute to its broadening.
In this sense, a better velocity dispersion determination can be derived from 
other forbidden lines. However, given the low metallicity of Irrs as a class,
the H$\alpha$ line is invariably adopted in FP studies (such as this one) 
due to its relatively high intensity.

Figure \ref{vr_maps} (top and bottom left) shows the global radial velocity
field of \objectname{NGC 4449} obtained from all the detected H$\alpha$
emission: \ion{H}{2} regions, bubbles, filaments, as well as the diffuse
emission.
Along the bar going to the SW, a conspicuous gradual decrement in radial 
velocities can be noticed ranging from approximately 250 km s$^{-1}$ in the 
NE to $\sim$ 190 km s$^{-1}$ in the SW. 
This result is in excellent agreement when it is compared against previous 
works \citep{crillon69, sabbadin84, malumuth86, yoshida95}. We are able to 
conclude that the gradient in radial velocities of the \ion{H}{2} gas is 
indeed in the opposite sense compared to the outer \ion{H}{1} distribution, 
as was reported in an earlier study by \citet{vanwoerden75} from \ion{H}{1} 
studies, and \citet{sabbadin84} from H$\alpha$ ones.

Figure~\ref{vr_maps} (bottom right) displays the global velocity dispersion
field (corrected as indicated in Section 4.2).
As can be seen, \objectname{NGC 4449} presents low velocity dispersion values
at its eastern and western sides, no larger than 25 km s$^{-1}$. However, the 
velocity dispersion values increases as one moves towards the interior of the 
galaxy, reaching 40 km s$^{-1}$.
The extreme values occur at the SNR location, as well as at some places along
the bar, with values in a range of 45--60 km s$^{-1}$. Our results along the
bar are in good agreement with the velocity dispersion map presented by
\citet{fuentes00}, and in disagreement with \citet{malumuth86}, who obtained
larger values (by about a factor of two) at the same locations. 
The latter authors may have precluded the signifance of a loss of parallelism
in their FP plates.

The recession velocity of the nucleus (which has been indicated with a cross
in Figure \ref{box-vel}) is 220 $\pm$ 4 km s$^{-1}$, in excellent
agreement according to \citet[223 $\pm$ 5 km s$^{-1}$]{sabbadin84} and
within the errors published in \citet[216 $\pm$ 28 km s$^{-1}$]{fouque92}.
With respect to our [\ion{S}{2}] data cubes, we generally find a
good agreement (where the signal/noise was high enough) with other H$\alpha$ 
radial velocity studies cited earlier in this paper.

\subsubsection{Local Superimposed Motions}

\begin{figure*}[htb]
\includegraphics{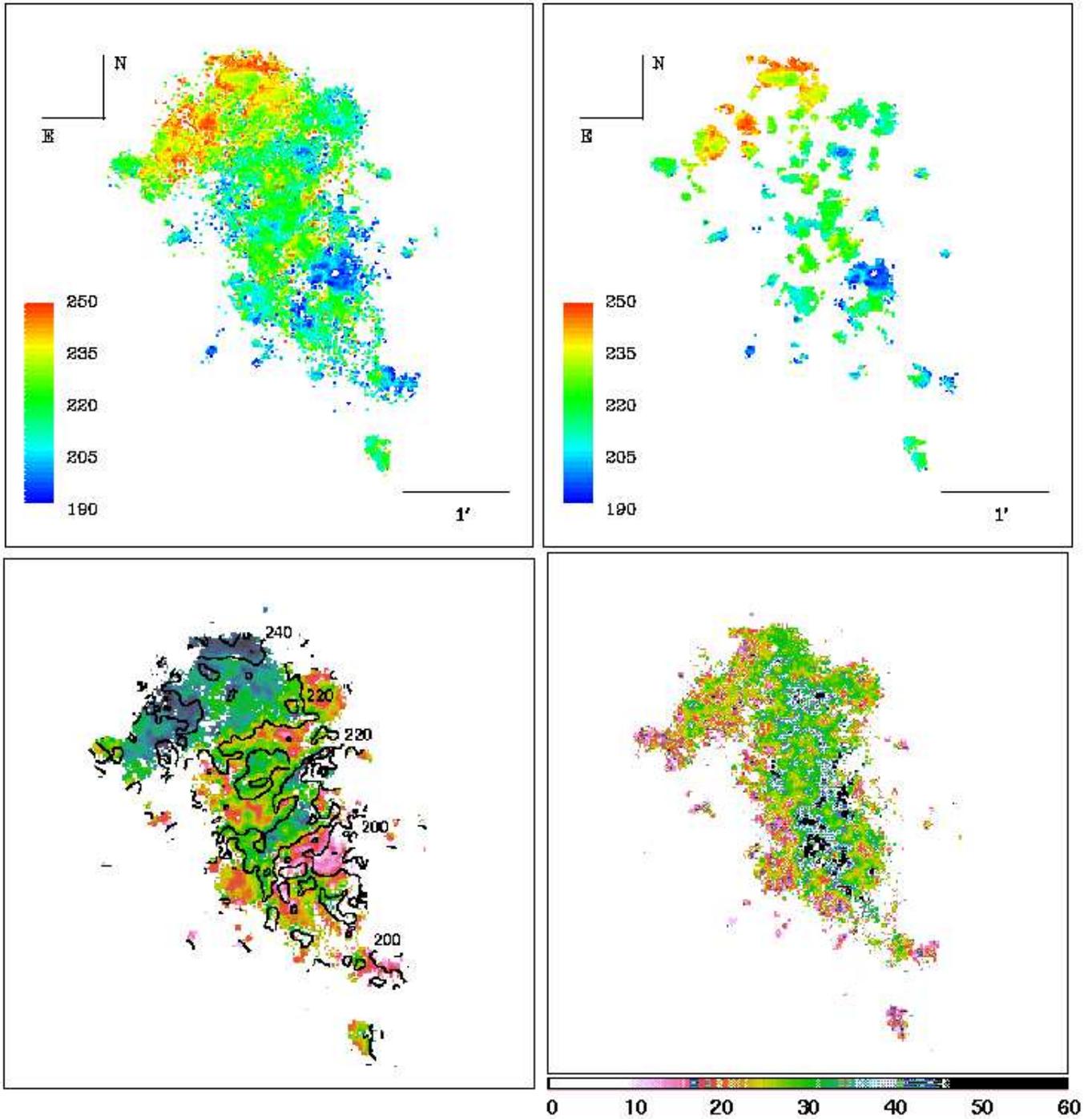}
\vspace{18.5cm}
\caption{
Top left: The observed heliocentric radial velocity field for the global
content of ionized gas (including the diffuse emission component). Top
right: As in the left panel, but for the ionized gas locally embedded 
in \ion{H}{2} regions.
Bottom left: Iso--velocities superimposed on the NGC 4449 radial
velocity field. Bottom right: A superposition of the velocity dispersion 
field. Both bottom panels pertain to the global content of ionized
gas in NGC 4449.
In the radial velocity fields we have selected a palette which better
enhances the gradient in velocity along the body of the galaxy 
(see text for details). The colors are related to heliocentric velocities. 
Velocity dispersions are given in \mbox{km s$^{-1}$}. The scale and the 
orientation are indicated.}
\label{vr_maps}
\end{figure*}

\begin{figure*}[htb]
\includegraphics{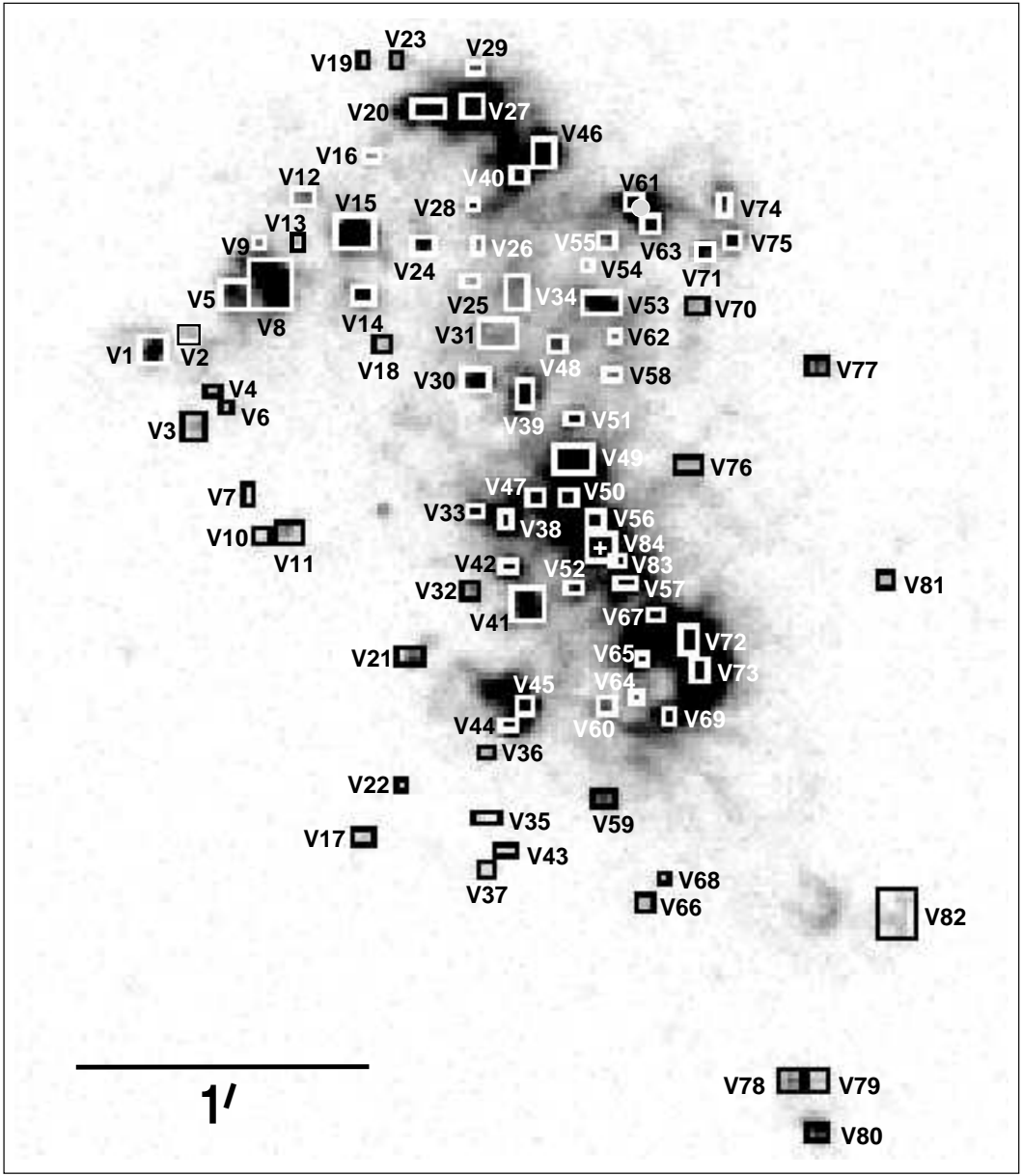}
\vspace{16.7cm}
\caption{Our H$\alpha$ monochromatic image of NGC 4449. The
diffuse component as well as the \ion{H}{2} region and filament 
populations, are readily noted.
The boxes correspond to locales where we performed our kinematical
analysis. The supernova remnant SB187 (embedded in our region V61) 
and the photometric center are indicated by a filled circle and by 
a cross, respectively.
The scale is indicated at bottom left. North is up; East to the left.}
\label{box-vel}
\end{figure*}

Besides the NE--SW velocity decrement in radial velocities, an 
intruiging substructure appears in our map: two ring--like structures 
(one of them located at the NW of the galaxy and embracing our regions 
V74 and V75; the other structure located at the north end of the bar 
enclosing our regions V18, V34, V49 and V53) are seen.
Such structures are noticeable in Figure \ref{vr_maps} (top 
and bottom left) and are comprised of both  \ion{H}{2} regions as well 
as diffuse emission.
When one studies the \ion{H}{2} regions alone --Figure~\ref{vr_maps} 
(top right)-- we note that such structures are not so apparent.
We actually find a spatial correspondence of the the second ring (described 
above) with a peculiar behavior of the magnetic vectors at this location. 
In Figure 3 of \citet{chyzy00}, a divergence in the polarization vectors 
occurs at low values (sometimes actually reaching zero), at this ring--like 
structure.
As the optical radial velocity field is somewhat chaotic at this locale,
it may be expected that magnetic fields do, to some degree, trace such gas 
kinematics (Beck and Urbanik, priv. comm.).

Furthermore, in Figure \ref{vr_maps} (top and bottom left) we find 
superbubbles, filaments and giant \ion{H}{2} complexes which have also been 
observed in previous studies \citep{hunter97, bomans97}. As far as the giant 
filament 6 \citep{hunter92} is concerned, we are able to only see it when 
spatial filtering was applied to our monochromatic chart.
More details of our analysis in this filament and in other filaments --as 
well as detailed results about the superbubble and SNR populations-- will be 
given in a forthcoming paper (Valdez--Guti\'errez et al. 2003, in preparation).

Extreme values in detected velocities (such as those arising from the internal 
motions of the superbubble, SNR and filament populations) may not properly be 
deduced in our Figure \ref{vr_maps} since we selected our palette to
enhance the velocity gradient along the major axis.
However, as discussed in the next section, when the kinematics of these
populations are analysed in detail, a complex and rather chaotic behavior 
emerges.
In our map, there are some regions where abrupt radial velocity gradients 
appear: for example, at the north end of the bar --where there are some clues 
that some superbubbles are blending-- and at the location of the superluminous 
SNR. Indeed, at the latter zone there is an \ion{H}{1} hole and the filaments 
5 and 6 reported by \citet{hunter92} lie at the borders of an \ion{H}{1} ridge.

In the surrounding velocity field outside of the bar, we find regions 
characterized by both large radial velocities and high velocity dispersions 
(details in the next section), in accord with \citet{malumuth86}.
Such locations are places where filaments and superbubbles have been reported
\citep{scowen92,bomans97,martin98}.

To explore this farther, we have analysed the radial velocity field at the 
seven locations where \citet{malumuth86} report strong radial velocity 
gradients (see their Figure 6 and our Figure~\ref{dig} top right).
After a detailed kinematical analysis --employing profile decompositions at
those locales--  our results agree qualitatively with Malumuth's and with our
values reported for the diffuse ionized component (see below).
We also find that the seven Malumuth's regions are coincident (at least in 
projection) with filaments or superbubbles. Due to the presence of such 
objects, the scenario proposed by \citet{hartmann86} in which gas cloud 
infall is generating such kinematics, could possibly be discarded. 
However, in order to elucidate the origin of the local, steep radial velocity 
gradients at the above mentioned locations, detailed kinematical studies have 
to be performed.
With respect to the velocity field at those locales where 
\citet[see their Figure 14]{hill98} postulated that star formation 
propagated in a stochastic manner, we find that the radial velocity 
profiles there are often complex, admitting more than one component 
(as can be seen in our Table~\ref{parameters_2}).
Such locations spatially correspond to regimes where \ion{H}{2} complexes,
filaments and bubbles are seen in projection.

We have already noted that the global velocity field in \objectname{NGC
4449} is moderately well behaved, but becomes increasingly chaotic as one 
considers smaller scales (as discussed in the the sections which follow).
Also, the projected borders of the bar exhibits abrupt gradients in radial 
velocity and in velocity dispersion.
What we therefore encountering here in the optical inner part of 
\objectname{NGC 4449} is a highly perturbed medium due to an intense star 
forming rate -- high compared to other Irrs \citep{hunter99}.
The star formation rate in this galaxy is undoubtedly enhanced by bar--like 
gravitational well.
We strongly believe that all of these kinematical and dynamical features 
suggest a prior encounter of this galaxy with the Irr \objectname{DDO 125}. 
This scenario was proposed by \citet{hunter98} based on \ion{H}{1} data 
covering a spatial extent of about 67$\arcmin\times$67$\arcmin$.
Later, \citet{theis01} performed numerical simulations which reinforced such 
a scenario.

\subsection{Kinematics of the \ion{H}{2} Regions in NGC 4449}

Figure \ref{box-vel} shows the H$\alpha$ monochromatic image of \objectname
{NGC 4449} (continuum subtracted) obtained from our FP cubes using the method
described in Section 2. Superimposed on this image are the
rectangular boxes (see Section 2.2) from which the radial velocity
measurements of the observed \ion{H}{2} regions have been computed.
We obtained the radial velocity profiles of these sources by integrating
within these rectangular boxes. We will refer to these profiles as 
``integrated'' because they are obtained by integrating over the whole extent 
of the \ion{H}{2} region. We also mark the position of the SNR with a filled 
circle.

In general, one single Gaussian function may be used for fitting the 
integrated radial velocity profiles. Only in five cases (V34, V56, V64, 
V65 and V76) are the integrated velocity profiles more complex. In these 
cases, we carried out a profile decomposition using two Gaussian functions. 
In these cases, we list the second velocity components (obtained from our 
profile fitting) in columns 4 and 6 of Table \ref{parameters_2}.

Table \ref{parameters_2} catalogues the kinematics of the \ion{H}{2} regions 
throughout the optical extent of \objectname{NGC 4449} spanned by our 
observations. We have labeled each region V1, V2, ... in column 1; cross 
identification of these regions in previously published papers are listed in 
column 2; the peak velocity is listed in columns 3 and 4; velocity dispersion 
values (from fitting a Gaussian to the radial velocity profiles) may be found 
in columns 5 and 6. Column 7 lists young stellar groups associated with our 
regions V1, V2, ... . while comments on the sources appear in column 8. 
In these comments, ``a" denotes asymmetric profiles, while ``s" demarcates 
symmetric ones.

The radial velocities were corrected to the heliocentric system of reference.
Our observational velocity dispersions ($\sigma_{obs}$ = FWHM$_{obs}$/2.354)
have been corrected for instrumental ($\sigma_{inst}$), thermal ($\sigma_{th}$)
and intrinsic ($\sigma_{intrin}$) broadenings, according to:

$$\sigma^2_{corr}= \sigma^2_{obs}-\sigma^2_{inst}-\sigma^2_{th}-
\sigma^2_{intrin}$$

\noindent where $\sigma_{inst}= 21.7$ km s$^{-1}$,
$\sigma_{th}= 9.1$ km s$^{-1}$ for hydrogen gas at T$_e$=10$^4$
\citep{spitzer78} and $\sigma_{intrin}=3$ km s$^{-1}$ for the fine structure
of the H$\alpha$ line. The resulting $\sigma_{corr}$ is also referred as the 
turbulent velocity dispersion by some authors.
The accuracy in the determination of peak velocities as well as error
determinations in the fitting of radial heliocentric velocities and velocity
dispersions related to PUMA data, have been previously reported in
\citet{rosado01}.

As can be seen in this Table, the \ion{H}{2} region heliocentric radial
velocities (of single profiles) range from 183 (source V22) to 254
km s$^{-1}$ (source V13). Within errors, these values are in agreement
with previously published values \citep{crillon69,sabbadin84,malumuth86}.
When compared to the study of \citet{fuentes00}, our radial velocities
show a tendency to be universally lower. By means of a least squares fit, 
the best slope value to Fuentes-Masip velocities is 0.74$\pm$0.07 with 
a correlation coefficient of 0.9; the amount of the difference reaches 
14 km s$^{-1}$ in the most extreme cases.

On the other hand, some extreme velocities were detected in regions with
composite profiles: their  velocity ranges from 185 (source V65) to 352 
km s$^{-1}$ (source V34). Inspecting our velocity charts, the objects 
related to these velocities are identified as \ion{H}{2} regions coinciding 
(in projection) with filaments discovered by \citet{sabbadin79}, 
\citet{scowen92}, \citet{hunter97}, \citet{bomans97} and \citet{martin98}.

The corrected velocity dispersions are supersonic for most of the objects,
with values ranging from 11.8 to 37.4 km s$^{-1}$ with a mean value of 22.8
km s$^{-1}$. This is in agreement with Fuentes-Masip et al. 2000 who
measured 44 \ion{H}{2} regions belonging exclusively to the bar.
Some relatively isolated \ion{H}{2} regions or \ion{H}{2} regions being 
part of a complex, had low S/N ratios and consequently the determinations of 
peak velocity and velocity dispersion were uncertain. Such cases are marked 
with a ``u"  in column 2 of Table \ref{parameters_2}.

Young stars related to our \ion{H}{2} regions are indicated in Column 7 of 
Table~\ref{parameters_2}. Such information is based on a previous study by 
\citet{hill98}. \citet{hill98} identified the locale of OB associations 
(as well as complexes of OB associations) using ratios of far ultraviolet 
(FUV) and optical imaging.
In that analysis,  OB associations and OB complexes were studied by
defining `small' and `large' apertures, respectively. Column 7 of Table 
\ref{parameters_2} indicates the young stellar content according to the 
work of \citet{hill98}: a number followed by a capital letter refers to a 
`large' aperture while stand--alone numbers indicates  the use of `small' 
apertures (see Tables 5 and 8 in Hill et al. 1988).
In that column, a blank space means that there is no indication of a 
young stellar population in Hill's data, or that some of our regions were 
not covered by their observations.
In general, the most notable complexes show a spatial correlation with young
stellar associations. Future studies with better resolution and sensitivity 
at both wavelengths (than those available to Hill et al. 1998) 
will undoubtedly 
provide further insight here.

\subsection{Kinematics of the Diffuse Ionized Gas in NGC 4449}

We have already noted that \objectname{NGC 4449} presents a diffuse, 
ionized gaseous component which permeates almost the entire optical 
extent of the galaxy. The nature of the diffuse ionized gas --also called 
DIG or warm ionized medium (WIM)-- which is found in the form of shells, 
loops, and a widespread unstructured diffuse background, has become a 
field of intense research. Review papers may be found in 
\citet{reynolds91,reynolds93}.

In view of the fact that our observations covered the optical disk of 
\objectname{NGC 4449}, we performed a detailed analysis of the properties 
of its extended DIG (i.e. the widespread diffuse background).
Contrary to what we find in the Local Group Irr \objectname{IC 1613}
\citep{valdez01}, the DIG in \objectname{NGC4449} is more turbulent
and reflects the influence of its activity at local scales (see below) and 
at global scales (see next section).

\begin{figure*}[htb]
\includegraphics{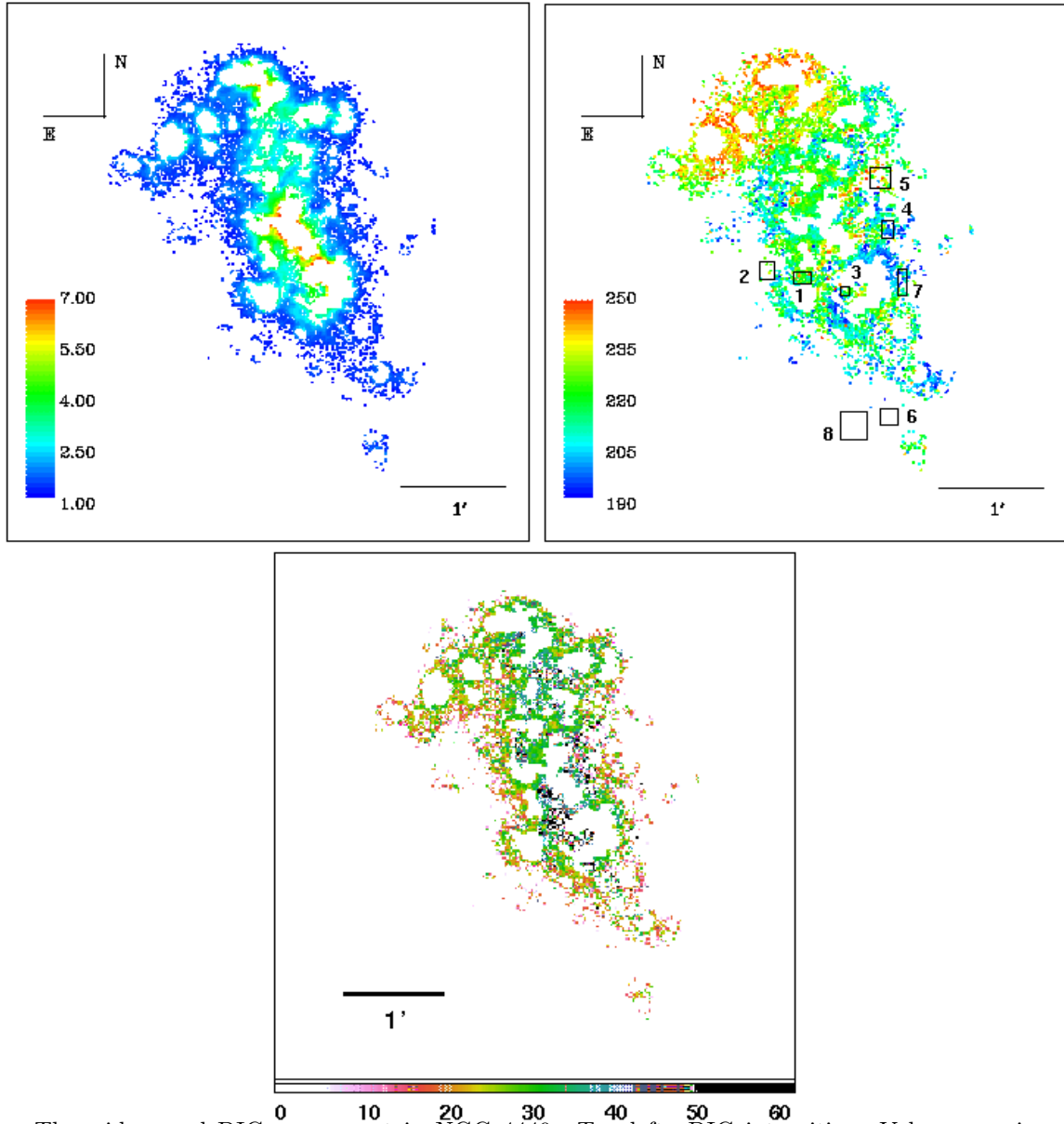}
\vspace{16.5cm}
\caption{The widespread DIG component in NGC 4449. 
Top left: DIG intensities. Values are given in counts s$^{-1}$, 
and are indicated in the respective palette. In order to determine 
H$\alpha$ fluxes, luminosities and surface brightnesses, multiplications 
by factors of 2.545$\times$10$^{-16}$ erg cm$^{-2}$
s$^{-1}$, 4.17$\times$10$^{35}$ erg s$^{-1}$ and 
3.165$\times$10$^{-5}$ erg cm$^{-2}$ s$^{-1}$ steradian$^{-1}$, 
respectively, are required.
Top right: The observed heliocentric radial velocity field. The boxes 
correspond to the locations of \citet{malumuth86}; see text for further
details. Bottom: The velocity dispersion field of NGC 4449. Values are 
given in km s$^{-1}$ in the latter two maps. Spatial orientation and the 
scale are indicated.}
\label{dig}
\end{figure*}

Concerning DIG photometric properties in \objectname{NGC 4449}, we 
find a gradual increment in surface brightness values as 
one moves from the outskirts of the galaxy towards the inner regions 
(see Figure~\ref{dig}, top left). 
In this way, surface brightness values at the outlying periphery 
of \objectname{NGC 4449} range from 3.165$\times$10$^{-5}$ to
2.215$\times$10$^{-4}$ erg cm$^{-2}$ s$^{-1}$ steradian$^{-1}$
(or 4.17$\times$10$^{35}$ -- 2.91$\times$10$^{36}$ erg s$^{-1}$ in
luminosity units) at the inner regions surrounding the brightest 
\ion{H}{2} complexes.
These values are approximately smaller by a factor of four than those 
found  by \citet[see their Figure 2]{fuentes00} in the diffuse gas 
at the bar locations.
However, our values found for the DIG in the outer extremities of 
the disk correspond well with those found by \citet{hunter90} at 
the locations of two loops along the bar.
The corresponding emission of the outlying DIG is $EM \sim 280$ 
cm$^{-6}$ pc, which falls in the range obtained by \citet{hunter90}.

We also compute the total DIG luminosity in \objectname{NGC 4449}. The 
DIG luminosity we find is 5.49$\times$10$^{39}$ erg sec$^{-1}$ (or
3.35$\times$10$^{-12}$ erg cm$^{-2}$ sec$^{-1}$), which is $\sim$10\% the
total H$\alpha$ luminosity reported by \citet{hunter99}.
Previous measurements of the DIG fraction in this galaxy \citep{hunter90}
report a range of 15--20\% of the total H$\alpha$ luminosity.
The disagreement with our value undoubtedly resides in the deeper
observations performed by Hunter and collaborators.
In the Large Magellanic Cloud, the DIG contribution is estimated to
be about 15--25\% of the total H$\alpha$ flux \citep{kennicutt86}.
These values are very small compared to the DIG luminosity in some spirals, 
which may amount to 30\% or even 50\% of the total H$\alpha$ luminosity
\citep{ferguson96}.

The kinematics of the DIG is displayed (according to the entire
spatial distribution) in Figure~\ref{dig} (top right). We show there 
the DIG's radial velocity field. To compute this, the \ion{H}{2} 
regions were subtracted from the whole field (the procedure was 
explained earlier, in Section 2.1).
In this map, a decreasing gradient of radial velocities from NE to SW
can be noted, which follows the tendency already seen in the total 
ionized gas content (\ion{H}{2} regions plus DIG; see previous sections), 
and in agreement with previous results (Section 4.1). Locally, part of 
the ring--like structure (discussed in Section 4.1.1), can barely be 
disentangled at the north end of the bar.
With respect to the velocity dispersion in the widespread DIG, we
find an increment in their values, from the peripheral optical domains
of \objectname{NGC 4449}, towards the interior of the disk.
The total displayed range in velocity dispersion is $\sim$70
km s$^{-1}$, varying from around 10 km s$^{-1}$ to $\sim$75 km
s$^{-1}$.
The maximum value we detected lay at the east of our region V65 (see
Figure~\ref{box-vel}). Moreover, values larger than 50 km s$^{-1}$ were 
also found east of the bar (between our regions V41, V42 and V52), as 
well as west of the nucleus. The latter locales lie close to our boxes 
V56, V83, V84; also at the NE sector of the bar (between our regions V33, 
V47 and V30), and at the vicinities of the SNR (surrounding our region V55).

\begin{figure*}[htb]
\includegraphics{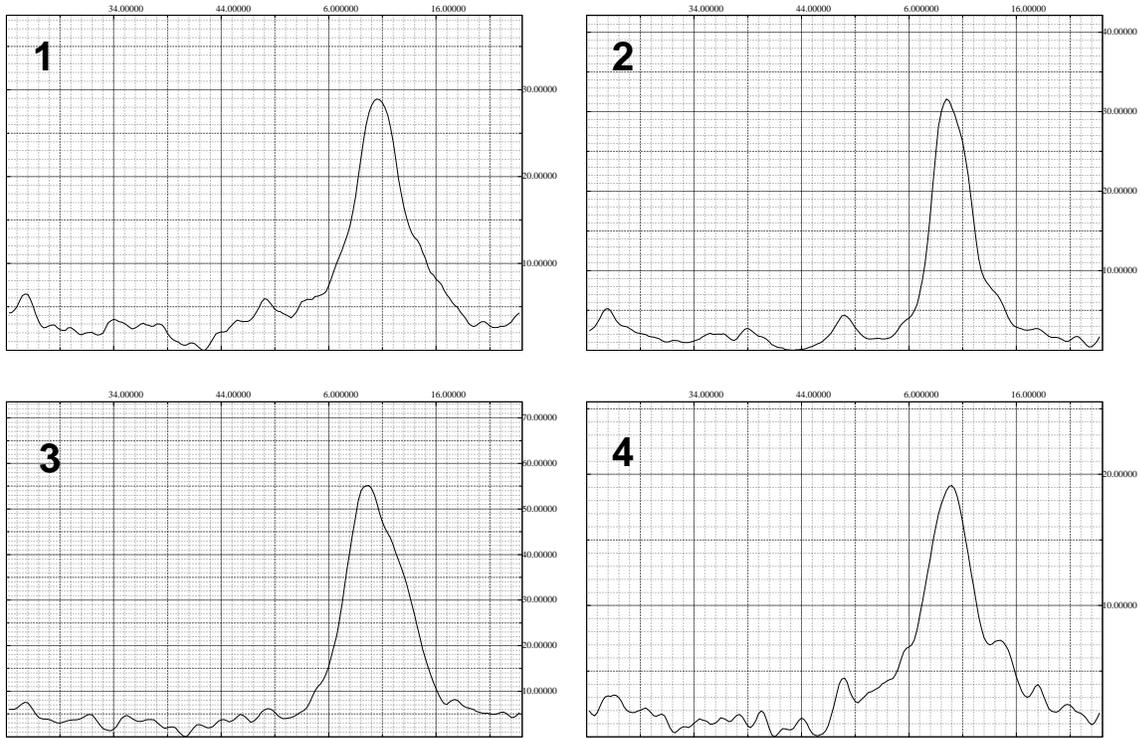}
\vspace{8.0cm}
\caption{Typical DIG profiles found in NGC 4449. The numbers in the
left hand corners correspond to the locales studied by \citet{malumuth86}.
Vertical and horizontal axes correspond to intensity (arbitrary units)
and to channel number, respectively. }
\label{malumuth_profiles}
\end{figure*}

In order to quantitatively analyse the kinematical properties in the
widespread DIG in \objectname{NGC 4449}, we performed a profile 
decomposition at a few locations.
The DIG radial velocity map displayed in Figure~\ref{dig} (top right),
was used to define boxes at different positions of interest. We then 
performed a profile decomposition within such boxes, as in section 4.2.
In this way, we could then secure kinematical properties of the DIG
in terms of heliocentric radial velocities and corrected velocity
dispersions.
The zones which displayed very peculiar profiles were those surrounding 
the bar, coinciding in many cases with regions where \citet{malumuth86} 
suggested a gas infall. Their positions are indicated in Figure~\ref{dig}.
These locales coincide --at least in projection-- with positions where 
filaments or shells have been reported 
\citep{hunter92,scowen92,bomans97,martin98}. Characteristic of
\objectname{NGC 4449} is its network of filaments, loops, shells, and 
the like. Pertaining to  the profile decomposition we performed at 
Malumuth's locations (see their Figure 6), we are confident they indeed 
are representative of the profiles at any location in the pervasive DIG 
of \objectname{NGC 4449}. Some examples of typical profiles we find at 
four of the \citet{malumuth86} locations are shown in 
Figure~\ref{malumuth_profiles}.

In general, the DIG profiles are very complex (multi--component fits
required per profile) at almost all locations. The main component 
in these profiles always retained values of $\pm$23 km s$^{-1}$ from 
the systemic velocity (220$\pm$4 km s$^{-1}$). Beside this main 
component, we find blue and  red components: the blue one having a 
range of 117 to 144 km s$^{-1}$ while the red component ranged from 
275 to 332 km s$^{-1}$.
With respect to the velocity dispersions, the DIG values in the main 
component reaches up to four times the velocity dispersion of the nearest
\ion{H}{2} region (see box 3, which is spatially in contact with
regions V60, V64 and V65).
The resulting DIG velocities and velocity dispersions were larger 
than the values we secured for the \ion{H}{2} regions closest to the
DIG boxes. We do however caution that fits to more than one component 
at such locations are to be treated carefully, given  high
perturbations in the velocity field (as already alluded to in our
previous discussions).

From our results of the global DIG maps (both in radial velocity and
in velocity dispersion; see Figure~\ref{dig}) as well as from our 
profile decomposition, we  find that some locations inside the widespread 
DIG in \objectname{NGC 4449} show a truly chaotic kinematical behavior.
It is tempting to postulate that some of the DIG locations we analysed 
have, as a source of their fueling, energetic photons escaping from 
\ion{H}{2} complexes.
Of course for the very widespread DIG over the disk of 
\objectname{NGC4449}, any inference with star forming regions and DIG 
is, at best, indirect.

Let us reflect back on the dynamical history of \objectname{NGC 4449}.
After a likely interaction with the Irr \objectname{DDO 125},
\citep{hunter98,theis01} a bar potential took form, in which gaseous
infall and star formation is taking place on a highly efficient basis
\citep{malumuth86}.
Other secondary manifestations (aftermaths of the interaction)
could be the loops, shells, filaments, etc: clear signatures of the
highly perturbed velocity field. Indeed, a past interaction experienced 
by \objectname{NGC 4449} could be the crucial factor as to why the
ISM in this galaxy is so active and peculiar.

\section{The Optical Rotation Curve of NGC 4449}

Given that we have kinematical data for the ionized gas spanning 
the entire optical disk, the next step is obviously to derive 
the rotation curve. However, we must point out that irregular 
galaxies invariably show low rotation velocities because of 
their low masses.

For \objectname{NGC 4449}, we calculate the rotation curve using the 
radial velocities resulting from our FP cubes. First, we use the 
first moment map or radial velocity chart, and we apply a mask. 
The mask is constructed using the monochromatic map previously
cleaned at 3$\sigma$ times the mean value of the background. In this
way, we obtain a radial velocity field with non--zero values in 
the positions where there is H$\alpha$ emission. We then calculate 
the rotation curve by using the classical formulae, as follows:

$$\Theta(R_k)={{v_k-v_0}\over{\cos {\theta}_k\sin i}}$$

\noindent where $\Theta(R_k)$ is the rotation velocity at the position 
$R_k$, $v_k$ is the heliocentric radial velocity at that position, $v_0$
is the systemic velocity, ${\theta}_k$ is the azimuthal position of the 
data point, and $i$ is the inclination of the galaxy \citep{mihalas81}.
This formulae assumes that the major axis is oriented at 
$\theta=0^{\circ}$.

We have developed a program which calculates the rotation curve once the 
central pixel, the systemic velocity, and the position and inclination 
angles (only points in $\pm 20^{\circ}$ around the major axis are
specified).
We compute the difference between the receding and approaching portions 
of the disk at any galactocentric radius.
While this method works well for spiral galaxies, caution should be
exercised in the case of irregulars, with their low rotation velocities 
and their intrinsic disk asymmetries.

For this galaxy, the dynamical center is assumed to lie at 
RA(2000)=12$^{\rm h} $28$^{\rm m}$14.7$^{\rm s}$ and
Dec(2000)=44$^{\circ}$05$^{'}$30$^{''}$ (which coincides with the 
photometric center). 
Contrary to what has been found for the \objectname{LMC} -- where
there is an offset between the optical center and the HI 
dynamical center \citep{kim98} --  in \objectname{NGC 4449},  
both optical and HI centres spatially coincide \citep{hunter99}.
We adopt $v_0=217$ km sec$^{-1}$, $PA=25^{\circ}$ and $i=43^{\circ}$.
Comparing the computed values with those in other papers, our $v_0$ is
in agreement with the value determined by \citet{bajaja94}.

Regarding the $PA$, our choice for this value disagrees with the
value given in the RC3 catalog \citep{devauco91}; We recall that
RC3 values are determined from optical Population I morphologies.
In a recent work \citep{hunter99}, it is noted that the $PA$ (in the 
Population I component) rotates E from N as the radius increases, 
indicative of the presence of a stellar bar potential.
Thus, the disagreement of our $PA$ with that in RC3 might conceivably 
be due to the fact that in this study we are exclusively dealing with 
the ionized gas component.
Finally, our inclination ($i$) agrees extremely well with \citet{tully88}.

\begin{figure*}[htb]
\includegraphics{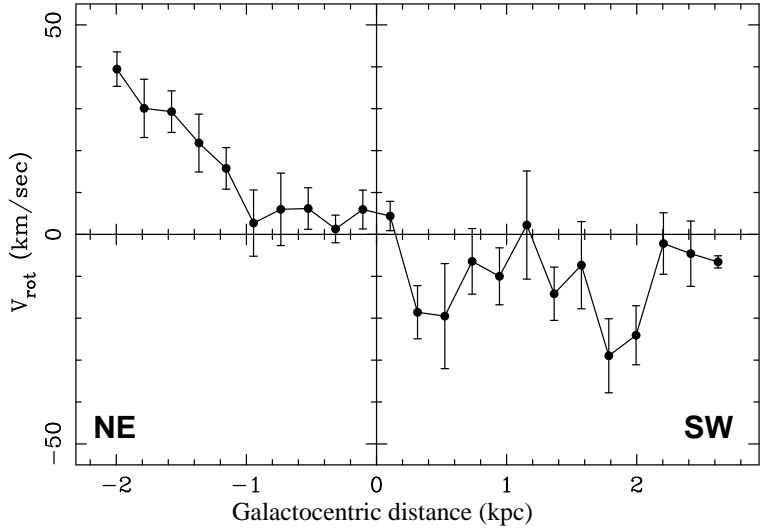}
\vspace{6.5cm}
\caption{Our H$\alpha$ rotation curve for NGC 4449. Full details
  appear in the
 main body of the paper.}
\label{rot_cur}
\end{figure*}

The resulting rotation curve is shown in Figure \ref{rot_cur}. From this
Figure we can see that the gas in this galaxy rotates slowly around the bar
perturbation with a maximum velocity value of $\sim$ 40 km s$^{-1}$ at a
radius of 2 kpc in the NE; on the contrary, the rotation curve displays
substantial velocity gradients at the SW zone.

It should be noted that the NE side of the galaxy (the receding side)
shows a smooth behavior which can be fitted by solid body rotation.
In contrast, the approaching side is less populated with \ion{H}{2} regions 
and the results are quite chaotic on local scales. 
Given the likely scenario of \objectname{NGC 4449} having undergone an 
interaction in the past \citep{hartmann86,hunter98,kohle99,theis99,theis01}, 
the highly perturbed radial velocity field in approaching side may bear 
some former signature of the interaction with \objectname{DDO 125} 
\citep{hunter98}.

When we compare our results with the \ion{H}{1} rotation curve calculated 
by \citet{martin98}, the agreement is good to within 30\% of the measured 
values (she reports 30 km s$^{-1}$ at 2 kpc).
From optical studies, \citet{malumuth86} report a {\it line--of--sight} 
rotation velocity of 15 km s$^{-1}$ in the bar. In this case, our velocity 
map in Figure 9 displays a {\it line--of--sight} rotation velocity of 
around $\sim$ 20 km s$^{-1}$. Taking into account our uncertainties --see
\citet{rosado01} -- the agreement is good.

Using our derived optical rotation curve, we calculate the mass inside 
the limiting radius reached by our observations. Taking the receding side 
in our rotation curve, the mass inside the optical radius of about 2 kpc, 
is 7.2 $\times$ 10$^8$ M$_\sun$. This result is in agreement with the 
optical mass derived in a previous work \citep{theis01}.
Using our derived mass value, we determine the central density in
\objectname{NGC 4449}; we derive a value  $\rho$= 0.021 M$_\sun$ pc$^{-3}$.
This value is in excellent agreement with the average value $\rho_c$ = 0.02
M$_\sun$ pc$^{-3}$, recently found for galactic halo central densities
\citep{firmani00,firmani01}. It is interesting to note that the central
halo densities are almost independent of galactic mass --whether dealing 
on galactic or galaxy cluster scales-- in frameworks of cold dark matter 
cosmologies.

In summary, we quantitatively find that the ionized gas in
\objectname{NGC 4449} is rotating weakly, as has been suggested
previously \citep{sabbadin84,hartmann86}. However, solid body rotation 
is only applicable for the NE sector of the disk. The SW side 
demonstrates abrupt radial velocity gradients, with differences greater 
than 100 km s$^{-1}$ for regions practically in contact (at least in 
projection). Due to substantial uncertainties in many of our data points 
at those SW locations, we did not deem it fit to determine the nature of 
the SW side rotation pattern in any further detail than explored here.

\section{Comparison Against the \ion{H}{1} Component}

\objectname{NGC 4449} has been recently studied in the \ion{H}{1} 
content by \citet{hunter99}. The spatial and velocity resolution
in that analysis was $9\farcs8\times 7\farcs78$ and 10 km s$^{-1}$, 
respectively. In our discussion here, we focus on the \ion{H}{1} 
component which spatially corresponds with our optical observations 
(Figures 14 and 16 in Hunter et al. 1999).

In the optical body of the galaxy, the \ion{H}{1} velocity field
follows the behavior which we determined for the ionized gas (compare 
our Figure~\ref{vr_maps} with Figure 14 in Hunter et al. 1999).
In the NE side of the optical body, the \ion{H}{1} velocity field 
presents values of the order of 240 km s$^{-1}$, while it decreases 
in the SW sector, to values of order 190 km s$^{-1}$.
As one approaches the outermost detected peripheries in the \ion{H}{1} 
distribution (inside the inner 13$\farcs$4), the velocity field 
is rather complex. Farther to the SW, the \ion{H}{1} velocity field 
increases in value (up to 350--400 km s$^{-1}$). 
This also includes the locale of the ``arc'' (west of the 
photometric center) already described in Section 3.
On the other hand, going farther to the NE, the \ion{H}{1} velocity 
field decreases to values of 150--190 km s$^{-1}$.
The outer \ion{H}{1} counterrotates relative to the inner \ion{H}{1}.
The complex behavior shown by the \ion{H}{1} velocity field is
probably a result of the interaction between \objectname{NGC 4449} and
\objectname{DDO 125} \citep{theis01}.

The \ion{H}{1} velocity dispersions inside the limits of the optical
disk are of the order of 15--35 km s$^{-1}$.  Along the optical bar,
the highest values (25--27 km s$^{-1}$) correspond to three locales,
as also found in this work and by \citet{malumuth86}.
Going to the NE end of the bar, the velocity dispersions experience 
an increment in their values, reaching 35 km s$^{-1}$ around the 
location where we report a ring--like feature (section 4.1.1), encircled 
by regions V25, V31, V34, and V48.
For the remainder of the northern sector of the disk, the velocity 
dispersions lie in the range of 12.5--20 km s$^{-1}$. In the optical 
disk, the \ion{H}{1} velocity dispersion field shows the same behavior 
we calculated with our FP data. The main difference is that the ionized 
gas presents higher velocity dispersions -- of order 2 times those 
in \ion{H}{1} (compare our Figure~\ref{vr_maps} with Figure 16 in 
Hunter et al. 1999).

This galaxy clearly does not present a classical picture of a
symmetric global velocity field in \ion{H}{1}; most likely again due
to the interloper \objectname{DDO 125} \citep{hunter98,theis01}.
The \ion{H}{1} column density distribution also betrays hints of a
past interaction. The peak of the \ion{H}{1} column density is located 
far away from the optical center, at a distance of approximately 
$50''$, which corresponds to 900 pc.
Also, as noted by \citet{hunter99}, there is little detailed
correlation between the column density of the \ion{H}{1} gas and the 
presence of \ion{H}{2} regions.

\section{Conclusions}

We have successfully performed a detailed kinematical and
dynamical analysis of the entire ionized gas content in the 
nearby irregular galaxy \objectname{NGC 4449}, by means of 
Fabry--Perot interferometry.
The analysis has been accomplished separately on both global 
as well as local scales and several of our conclusions
are new.

On local scales, we focused our attention on the \ion{H}{2} region
population, extracting radial velocities and velocity dispersions.
In this way, we present, for the first time, the most extensive 
kinematical catalog of the \ion{H}{2} regions in this galaxy 
yet published.
We find that the \ion{H}{2} region population shows similar
integrated properties when compared to kinematical data
previously obtained in the relatively isolated irregular 
\objectname{IC 1613} \citep{valdez01}.
The latter statement also applies to the velocity dispersions 
in the star forming regions: in both galaxies, they are approximately 
identical.
Moreover, in many cases we did detected velocity profiles
indicative of a complex structure not fitted by {\it single} gaussian 
curves. These profiles are undoubtedly related to the high perturbed 
radial velocity field, which shows traces of a plethora of
superbubbles, filaments and a very active (and widespread) diffuse
ionized gas spanning the entire optical disk.
Statistically, using the \ion{H}{2} region diameter distribution and
the \ion{H}{2} region luminosity function, we find that the star
forming complexes are typical of those obtained for larger samples of 
irregulars.

We furthermore analysed the kinematical properties of the diffuse 
ionized gas (DIG) on global scales by means of radial velocity and velocity 
dispersion fields; locally, we explored the kinematics, making use of
profile decomposition. 
As far as we are aware, this is the first time that such an
analysis has been performed in this object.
Our results point towards a highly chaotic medium, especially at 
those locations close to the brightest \ion{H}{2} complexes in the bar.
We postulate that the chaotic radial velocity field induces local 
compression ensuring the propagation of stochastic star formation
--as also discussed by \citet{hartmann86}.

On global scales, we find a relatively well behaved radial velocity
field, showing a decreasing gradient in radial velocity along the optical
bar from NE to SW, suggesting solid body rotation. This result agrees 
well with previous kinematical results \citep[and references therein]{valdez00}.
We believe that \objectname{NGC 4449} presents solid body rotation 
(albeit at low rotation velocities) north of the optical center.
Reliable fits are not possible in the southern sector of the disk, due
to the chaotic motions at those locales.
The inner \ion{H}{1} kinematics (inner 13$\farcm$4) also shows 
a good correspondence with the optical radial velocities we find in 
this study.

In general, our kinematical and dynamical findings for the ionised
gas in \objectname{NGC 4449} are fully consistent within the framework 
of a past interaction with \objectname{DDO 125}.

\acknowledgements{
MVG acknowledges support from a CONACyT scholarship number 114735.
This work was partially supported by grants IN122298 of 
DGAPA--UNAM, 27984--E and 28507--E of CONACYT.
The authors wish to thank Drs. D. Mayya, A. Luna and E. Brinks, for 
their comments and criticism pertinent to our study. Fruitful 
discussions with Drs. M. Bureau, I. Shlosman and V. Avila--Reese, 
enriched the last phase of this work and took place at the 
International Workshop in Astrophysics ``Guillermo Haro 2001".
The authors are indebted to Prof. David Block for kindly
assisting us with the grammatical content of our paper, and for 
discussions which greatly improved the scientific presentation 
of the contents. 
MVG thanks P. Scowen for providing her with his PhD Thesis, which
was useful when performing comparisons with the present work. IP thanks 
J. Boulesteix and H. Plana for providing advice on the Adhoc package.
We also thanks the anonymous referee for her/his insight and 
suggestions.
This research has made use of the NASA/IPAC Extragalactic Database (NED)
which is operated by the Jet Propulsion Laboratory, California Institute
of Technology, under contract with the National Aeronautics and Space
Administration.}

\begin{deluxetable}{lll}
\tablewidth{0pc}
\tablecaption{NGC 4449 global properties and observed
quantities\label{parameters}}
\tablehead{ \colhead{Parameter}  & \colhead{Value}  & \colhead{Reference} \\ }
\startdata
Name                         & NGC 4449  &                 \\
Type\tablenotemark{a}        & I(B)m     & \citet{tully88} \\
R.A. (2000)\tablenotemark{b} & 12$^{\mathrm h}$ 28$^{\mathrm m}$ 13$\fs$1 &
\citet{devauco76} \\
Dec (2000)\tablenotemark{b}  & +44$\arcdeg$ 05$\arcmin$ 43$\arcsec$      \\
Distance\tablenotemark{c}          & 3.7 Mpc           & \citet{bajaja94} \\
B magnitude\tablenotemark{d}       & --18.5            & \citet{schmidt92} \\
Angular size\tablenotemark{e}      & 5$\farcm$8$\times$4$\farcm$5 & LEDA   \\
Radial velocity\tablenotemark{f}   & 211 km s$^{-1}$    & NED     \\
Axial ratio                        & 0.77               & \citet{tully88} \\
Inclination                        & 43\degr            & \citet{tully88} \\
P.A.\tablenotemark{g}               & 45\degr            & \citet{devauco91} \\
M$_{tot}$\tablenotemark{c} & 7$\times$10$^{10}$ M$_\sun$ & \citet{bajaja94}\\
M$_{\rm HI}$\tablenotemark{c}  & 2.3$\times$10$^9$ M$_\sun$ &\citet{bajaja94} \\
HI extent\tablenotemark{c} & 43$\times$70 kpc           &\citet{bajaja94} \\
L$_B$\tablenotemark{h}     & 2.1$\times$10$^9$ L$^B$$_\sun$& \citet{martin98}\\
(M/L)$_*$\tablenotemark{h} & 11                         &  \citet{martin98}\\
SFR\tablenotemark{i}       & 0.47 M$_\sun$ yr$^{-1}$    &\citet{hunter99} \\
Log L(H$\alpha$)\tablenotemark{i} & 40.82 erg s$^{-1}$  &\citet{hunter99} \\
\enddata
\tablenotetext{a}{Morphologycal type.}
\tablenotetext{b}{Photometric center position.}
\tablenotetext{c}{Distance, total mass, \ion{H}{1} mass and \ion{H}{1} extent
adopted in this work.}
\tablenotetext{d}{Absolute B magnitude.}
\tablenotetext{e}{Optical extent. Isophotal diameter at 25$^m$ per
arcsecond squared.}
\tablenotetext{f}{Optical heliocentric radial velocity. The error in this
measurement is $\pm$ 15 km s$^{-1}$.}
\tablenotetext{g}{Position angle measured on the stellar component.}
\tablenotetext{h}{B-band luminosity (where M$^B$$_\sun$=5.48) and
stellar mass to blue light ratio in absence of dark matter.}
\tablenotetext{i}{Star formation rate and log of the H$\alpha$ luminosity.}
\end{deluxetable}

\begin{deluxetable}{lccccc}
\tablewidth{0pc}
\tablecaption{PUMA observations of NGC 4449 \label{observations}}
\tablehead{
\colhead{Mode\tablenotemark{a}} & \colhead{Filter}                &
\colhead{bin\tablenotemark{b}}  & \colhead{t\tablenotemark{c}}    & \colhead{Date}
& \colhead{Comments} \\ }
\startdata
FP & H$\alpha$ & 2 & 5760&  09.01.1997 & 2 datacubes\\
FP & [SII]     & 2 & 5760&  20.06.1998 & 2 datacubes\\
D  & H$\alpha$ & 2 & 120 &  10.01.1997 &  \\
D  & [SII]     & 2 & 120 &  10.01.1997 &  \\
D  & [OIII]    & 2 & 300 &  10.01.1997 &  \\
D  & [NII]     & 2 & 120 &  10.01.1997 &  \\
D  & H$\alpha$ & 1 & 60  &  20.06.1998 &  \\
D  & [SII]     & 1 & 120 &  20.06.1998 &  \\
D  & [OIII]    & 1 & 120 &  20.06.1998 &  \\
D  & [NII]     & 1 & 120 &  20.06.1998 &  \\
D  & H$\alpha$ & 1 & 30  &  20.06.1998 & Standard star \\
D  & H$\alpha$ & 1 & 30  &  20.06.1998 & Standard star\\
D  & H$\alpha$ & 4 & 60  &  11.02.1999 & Position 1 \\
D  & H$\alpha$ & 4 & 300 &  11.02.1999 &  \\
D  & H$\alpha$ & 4 & 60  &  11.02.1999 & Position 2 \\
D  & H$\alpha$ & 4 & 300 &  11.02.1999 &  \\
D  & H$\alpha$ & 4 & 60  &  11.02.1999 & Position 3 \\
D  & H$\alpha$ & 4 & 300 &  11.02.1999 &  \\
\enddata
\tablenotetext{a}{Mode of observation: FP= scanning Fabry--Perot observations,
D= direct imaging.}
\tablenotetext{b}{Binning.}
\tablenotetext{c}{Total integration time in seconds.}
\end{deluxetable}

\begin{deluxetable}{llcccc}
\tablewidth{0pc}
\tabletypesize{\footnotesize}
\tablecaption{Photometry for the \ion{H}{2} regions in NGC 4449\label
{parameters_1} }
\tablehead{
\colhead{ID} & \colhead{Other IDs} & \colhead{D} & \colhead{log F(H$\alpha$)}
& \colhead{log L(H$\alpha$)}       & \colhead{log S(H$\alpha$)}
\\
\colhead{} & \colhead{} & \colhead{(pc)} & \colhead{erg cm$^{-2}$ s$^{-1}$} &
\colhead{erg s$^{-1}$} & \colhead{erg cm$^{-2}$ s$^{-1}$ steradian$^{-1}$}
}
\startdata
M1  & 3            &  59.1 & -13.91 & 37.31 & -4.22 \\
M2  & 4            &  68.1 & -13.67 & 37.54 & -4.11 \\
M3  & 6,7, V1      & 129.9 & -12.78 & 38.44 & -3.78 \\
M4  & 8, V2        &  49.0 & -13.91 & 37.30 & -4.07 \\
M5  & 9, V3        &  67.7 & -13.58 & 37.63 & -4.01 \\
M6  & 12, V4       &  52.1 & -13.94 & 37.28 & -4.14 \\
M7  & 13,14,16, V5 &  99.4 & -12.96 & 38.25 & -3.73 \\
M8  & 15, V6       &  66.4 & -13.78 & 37.42 & -4.20 \\
M9  & 17           &  48.4 & -13.61 & 37.60 & -3.75 \\
M10 & 18,19, V7    &  39.1 & -14.33 & 36.89 & -4.28 \\
M11 & 20, V8       &  67.9 & -13.09 & 38.12 & -3.53 \\
M12 & 21, V9       &  57.1 & -13.60 & 37.62 & -3.88 \\
M13 & 22, V10      &  51.1 & -14.03 & 37.18 & -4.22 \\
M14 & 23, V8       &  76.1 & -12.95 & 38.26 & -3.49 \\
M15 & 24, V8       & 103.0 & -12.61 & 38.60 & -3.41 \\
M16 & 26,27, V11   &  80.7 & -13.43 & 37.78 & -4.02 \\
M17 & 28,31, V12   &  68.2 & -13.46 & 37.75 & -3.90 \\
M18 &34,35,37,38,39,41,43, V15
                     & 183.3 & -12.21 & 39.00 & -3.51 \\
M19 & 36             &  45.3 & -13.85 & 37.36 & -3.94 \\
M20 & 40, V16        &  49.4 & -13.78 & 37.44 & -3.94 \\
M21 & 44, V17        & 105.6 & -13.47 & 37.74 & -4.29 \\
M22 & 45             &  34.7 & -14.17 & 37.05 & -4.02 \\
M23 & 47, V18        &  94.1 & -12.91 & 38.31 & -3.63 \\
M24 & 48,54, V19,V23 &  89.7 & -13.38 & 37.84 & -4.06 \\
M25 & 50,55,59, V20  & 164.5 & -12.13 & 39.09 & -3.33 \\
M26 & 51,58, V21     &  87.6 & -13.27 & 37.94 & -3.93 \\
M27 & 52,56,57, V24   &  83.2 & -13.12 & 38.09 & -3.73 \\
M28 & 53, V22         &  91.0 & -13.80 & 37.41 & -4.49 \\
M29 & 62,64           &  83.2 & -13.59 & 37.63 & -4.20 \\
M30 & 63              &  71.8 & -13.65 & 37.56 & -4.14 \\
M31 & 73              &  63.6 & -13.99 & 37.23 & -4.37 \\
M32 & 75, V27         & 184.4 & -11.68 & 39.53 & -2.98 \\
M33 & 76, V28         &  63.5 & -13.33 & 37.88 & -3.71 \\
M34 & 78              &  82.0 & -13.44 & 37.78 & -4.04 \\
M35 & 79, V29         &  83.3 & -13.21 & 38.00 & -3.83 \\
M36 & 81,82, V30      & 127.5 & -12.66 & 38.55 & -3.65 \\
M37 & 83              &  71.2 & -13.67 & 37.55 & -4.14 \\
M38 & 84, V31         &  66.6 & -13.50 & 37.72 & -3.92 \\
M39 & 86, V32         &  57.8 & -13.66 & 37.56 & -3.95 \\
M40 & 89, V33         & 104.5 & -12.81 & 38.40 & -3.61 \\
M41 & 90, V32         &  38.8 & -14.17 & 37.04 & -4.12 \\
M42 & 91              &  60.3 & -13.07 & 38.15 & -3.40 \\
M43 & 92,103          &  75.5 & -13.24 & 37.98 & -3.77 \\
M44 & 93,107,108, V34 & 138.3 & -12.84 & 38.38 & -3.89 \\
M45 & 94,99,116       & 134.4 & -12.60 & 38.61 & -3.63 \\
M46 & 96, V35          &  82.2 & -13.72 & 37.50 & -4.32 \\
M47 & 98, 105, V36     &  77.6 & -13.58 & 37.63 & -4.13 \\
M48 & 100, V37         &  90.1 & -13.63 & 37.59 & -4.31 \\
M49 & 108, V38         & 130.4 & -12.35 & 38.86 & -3.35 \\
M50 & 111, V39         &  64.4 & -13.35 & 37.87 & -3.74 \\
M51 & 112, V40         & 117.9 & -12.56 & 38.66 & -3.47 \\
M52 & 113,115, V41     &  71.6 & -13.43 & 37.78 & -3.92 \\
M53 & 114, V42         & 115.7 & -12.72 & 38.50 & -3.62 \\
M54 & 117,121, V39     & 109.5 & -12.73 & 38.48 & -3.58 \\
M55 & 118, V43         &  66.0 & -13.87 & 37.34 & -4.28 \\
M56 & 119,123, V44     &  87.2 & -12.78 & 38.44 & -3.43 \\
M57 & 125, V45         &  94.5 & -12.84 & 38.38 & -3.56 \\
M58 & 126,127, V46     & 151.4 & -12.30 & 38.91 & -3.43 \\
M59 & 128,129, V41     &  91.0 & -12.94 & 38.27 & -3.64 \\
M60 & 131,137, V41     &  67.1 & -13.22 & 38.00 & -3.64 \\
M61 & 132,134, V45     &  97.8 & -12.62 & 38.59 & -3.37 \\
M62 & 133, V47         & 112.9 & -12.53 & 38.67 & -3.41 \\
M63 & 138,146,158, V49 & 123.8 & -12.48 & 38.73 & -3.44 \\
M64 & 139,140, V48     &  72.4 & -13.30 & 37.92 & -3.79 \\
M65 & 142,150,151,157, V50
                       & 136.0 & -12.33 & 38.89 & -3.37 \\
M66 & 152, V51         &  92.5 & -13.03 & 38.19 & -3.73 \\
M67 & 153,159, V49     &  86.4 & -12.90 & 38.31 & -3.54 \\
M68 & 161, V52         &  85.0 & -13.08 & 38.13 & -3.71 \\
M69 & 165,169,172,186,188, V53
                  & 128.6 & -12.66 & 38.56 & -3.65 \\
M70 & 167         &  54.8 & -13.60 & 37.61 & -3.85 \\
M71 & 168, V55    &  64.8 & -13.40 & 37.81 & -3.80 \\
M72 & 170,171, V56   &  70.9 & -12.87 & 38.34 & -3.34  \\
M73 & 173         &  51.4 & -13.79 & 37.43 & -3.98 \\
M74 & 180, V58         &  77.9 & -13.31 & 37.90 & -3.87 \\
M75 & 181,182, V59     &  98.5 & -13.17 & 38.05 & -3.93 \\
M76 & 183, V60         &  87.5 & -13.20 & 38.00 & -3.86 \\
M77 & 185,187, V61     & 142.6 & -12.43 & 38.79 & -3.51 \\
M78 & 191         &  70.6 & -13.07 & 38.15 & -3.54 \\
M79 & 192         & 100.7 & -12.64 & 38.58 & -3.41 \\
M80 & 199,200     &  51.1 & -13.58 & 37.63 & -3.77 \\
M81 & 202         &  93.9 & -13.21 & 38.01 & -3.93 \\
M82 & 203, V65         &  84.7 & -12.85 & 38.35 & -3.48 \\
M83 & 204, V66         &  81.7 & -13.60 & 37.61 & -4.20 \\
M84 & 208,209, V67     &  89.5 & -12.70 & 38.50 & -3.38 \\
M85 & 212, V69         & 107.4 & -12.66 & 38.55 & -3.49 \\
M86 & 218,219, V71     & 111.4 & -12.98 & 38.24 & -3.84 \\
M87 & 220, V72         &  95.9 & -12.29 & 38.91 & -3.03 \\
M88 & 221,222, V73     &  78.7 & -12.50 & 38.70 & -3.07 \\
M89 & 224,225,227, V74 & 132.0 & -12.85 & 38.37 & -3.86 \\
M90 & 226,234, V73     &  78.2 & -12.51 & 38.70 & -3.07 \\
M91 & 230,231, V75     & 108.0 & -12.86 & 38.36 & -3.70 \\
M92 & 233,235, V76     &  56.1 & -13.93 & 37.28 & -4.20 \\
M93 & 237,238, V77     &  98.4 & -13.24 & 37.98 & -3.99 \\
M94 & 239,240,244, V78, V79 & 149.9 & -12.99 & 38.22 & -4.12 \\
M95 & 245,247,248, V80 & 151.2 & -12.87 & 38.35 & -4.00 \\
M96 & 249, V81         &  67.8 & -13.71 & 37.50 & -4.15 \\
M97 & 250,252, V82     & 127.6 & -13.25 & 37.96 & -4.23 \\
M98 & 251, V82         &  81.7 & -13.63 & 37.58 & -4.23 \\
M99 & nucleus, V84     & 163.0 & -11.89 & 39.32 & -3.07 \\
M100 & h61        &  67.0 & -12.84 & 38.37 & -3.26 \\
M101 & h66        &  69.5 & -12.45 & 38.75 & -2.91 \\
\enddata
\end{deluxetable}

\begin{deluxetable}{llllllcl}
\tablewidth{0pc}
\tabletypesize{\footnotesize}
\tablecaption{Measured kinematical parameters for the \ion{H}{2} regions in
NGC 4449\label
{parameters_2} }
\tablehead{
\colhead{ID}         & \colhead{Other IDs}       &
\colhead{$v_1$}      & \colhead{$v_2$}           & \colhead{$\sigma_1$} &
\colhead{$\sigma_2$} & \colhead{Stars} & \colhead{comments}
\\
\colhead{} & \colhead{} & \colhead{(km s$^{-1}$)} & \colhead{(km s$^{-1}$)} &
\colhead{(km s$^{-1}$)} & \colhead{(km s$^{-1}$)} & \colhead{}
}
\startdata
V1  & 6            & 218 &\nodata& 17.6 &\nodata&  1N, 28  & a \\
V2  & 8u           & 230 &\nodata& 21.2 &\nodata&  1N,\nodata & a, dig contaminated? \\
V3  & 9            & 220 &\nodata& 18.5 &\nodata& \nodata & s \\
V4  & 12           & 227 &\nodata& 17.6 &\nodata& \nodata & a \\
V5  & 13,14,16     & 231 &\nodata& 22.1 &\nodata&  2N, 23 & a \\
V6  & 15           & 228 &\nodata& 22.9 &\nodata& \nodata & a \\
V7  & 18u,19u      & 220 &\nodata&  4.3 &\nodata&  3O,\nodata & a \\
V8  & 20,23,24     & 236 &\nodata& 24.7 &\nodata&  2N, 24  & a \\
V9  & 21u          & 243 &\nodata& 27.3 &\nodata&  2N,\nodata & a, wide wings, dig?\\
V10 & 22           & 211 &\nodata& 17.6 &\nodata&  3O,\nodata & a \\
V11 & 26,27        & 201 &\nodata& 19.5 &\nodata&  3O, 35  & a \\
V12 & 28,31        & 244 &\nodata& 26.4 &\nodata& \nodata, 8  & s \\
V13 & 29           & 254 &\nodata& 11.8 &\nodata& \nodata & a \\
V14 & 33u,36       & 231 &\nodata& 23.9 &\nodata& \nodata & s \\
V15 &34,35,37,38,39,41,43
                   & 244 &\nodata& 23.9 &\nodata&  4N, 13 & s \\
V16 & 40           & 231 &\nodata& 26.4 &\nodata& \nodata & s \\
V17 & 44           & 197 &\nodata& 14.8 &\nodata& \nodata & a \\
V18 & 47           & 211 &\nodata& 22.1 &\nodata& \nodata & a, red broad wing \\
V19 & 48           & 246 &\nodata& 15.1 &\nodata& \nodata & a \\
V20 & 50,55,59     & 235 &\nodata& 21.2 &\nodata&  5N, 2  & a \\
V21 & 51,58,63u    & 207 &\nodata& 19.5 &\nodata& \nodata, 44 & a \\
V22 & 53           & 183 &\nodata& 17.6 &\nodata& \nodata & a \\
V23 & 54           & 243 &\nodata& 22.1 &\nodata& \nodata & a \\
V24 & 56,57        & 228 &\nodata& 20.4 &\nodata& \nodata, 15 & a \\
V25 & 72           & 232 &\nodata& 27.3 &\nodata& \nodata & a, broad red wing \\
V26 & 74,80        & 230 &\nodata& 22.1 &\nodata& \nodata & s \\
V27 & 75,91        & 231 &\nodata& 22.1 &\nodata&  6N, 1 & a \\
V28 & 76           & 236 &\nodata& 22.1 &\nodata&  9N,\nodata & a \\
V29 & 79           & 246 &\nodata& 28.2 &\nodata&  6N,\nodata & a \\
V30 & 81,82        & 211 &\nodata& 17.6 &\nodata& 10B, 20 & a \\
V31 & 84,85,97     & 216 &\nodata& 33.2 &\nodata& \nodata & a \\
V32 & 86,90u       & 220 &\nodata& 17.6 &\nodata& \nodata & s \\
V33 & 89           & 207 &\nodata& 19.5 &\nodata& 11B, 33 & a \\
V34 & 93,107,108   & 216 & 352   & 26.2 & 12.9  & \nodata & double \\
V35 & 96           & 205 &\nodata& 14.5 &\nodata& \nodata & a \\
V36 & 98, 105      & 196 &\nodata& 20.4 &\nodata&  8B,\nodata & a \\
V37 & 100          & 192 &\nodata& 17.6 &\nodata& \nodata & a \\
V38 & 108          & 218 &\nodata& 19.5 &\nodata& 11B, 34 & a \\
V39 & 111,117,121  & 226 &\nodata& 29.0 &\nodata& 10B, 30 & s \\
V40 & 112          & 242 &\nodata& 24.7 &\nodata&  9N, 6 & a \\
V41 & 113,115,128,129,131,137
                   & 223 &\nodata& 20.4 &\nodata& 12B, 39 & a, wide wings \\
V42 & 114          & 221 &\nodata& 17.6 &\nodata& 12B, 38 & a \\
V43 & 118          & 205 &\nodata& 20.4 &\nodata& \nodata & a \\
V44 & 119,123      & 205 &\nodata& 18.5 &\nodata&  8B, 50 & a \\
V45 & 125u,132     & 216 &\nodata& 18.5 &\nodata&  8B, 49 & a \\
V46 & 126,127      & 236 &\nodata& 26.4 &\nodata& 13N, 4  & s \\
V47 & 133          & 207 &\nodata& 22.1 &\nodata& 11B, 32 & a \\
V48 & 139,140      & 219 &\nodata& 26.4 &\nodata& \nodata & s, broad red wing \\
V49 & 146,147,153,158,159
                   & 222 &\nodata& 29.0 &\nodata& 14B, 31 & a \\
V50 & 151          & 219 &\nodata& 30.7 &\nodata& 14B,\nodata & a \\
V51 & 152          & 219 &\nodata& 22.1 &\nodata& \nodata & a, wide wings \\
V52 & 161          & 237 &\nodata& 35.7 &\nodata& 12B,\nodata & s, wide wings \\
V53 & 165,169,172  & 203 &\nodata& 23.4 &\nodata& 15N, 22 & a \\
V54 & 166          & 227 &\nodata& 32.4 &\nodata& \nodata, 18 & a, dig contaminated?\\
V55 & 168          & 223 &\nodata& 37.4 &\nodata& \nodata, 11 & a \\
V56 & 170          & 192 & 256   & 20.4 & 19.5  & 16B, 37 & double \\
V57 & 178,190      & 219 &\nodata& 29.0 &\nodata& 16B, 37 & a \\
V58 & 180          & 223 &\nodata& 36.5 &\nodata& \nodata & s, top flated \\
V59 & 181,182      & 215 &\nodata& 22.1 &\nodata& \nodata, 51 & a \\
V60 & 183          & 189 &\nodata& 26.4 &\nodata& 18B,\nodata& a, broad red wing  \\
V61 & 185,187      & 220 &\nodata& 26.4 &\nodata& 17N, 9 & a, double?, SNR \\
V62 & 186          & 205 &\nodata& 30.7 &\nodata& \nodata & a, broad red wing \\
V63 & 192          & 218 &\nodata& 26.4 &\nodata& 17N, 10  & a, SNR     \\
V64 & 197          & 188 & 253   & 20.4 & 20.4  & 18B, 45  & double \\
V65 & 203          & 185 & 255.0 & 19.5 & 17.6  & 18B, 42  & double\\
V66 & 204          & 207 &\nodata& 20.4 &\nodata& \nodata  & a, broad blue wing\\
V67 & 208,209      & 197 &\nodata& 26.4 &\nodata& 18B, 40 & a \\
V68 & 211          & 218 &\nodata& 19.5 &\nodata& \nodata & a \\
V69 & 212          & 205 &\nodata& 23.8 &\nodata& 18B, 46 & a \\
V70 & 213,217      & 222 &\nodata& 29.0 &\nodata& \nodata, 21  & s \\
V71 & 218,219      & 220 &\nodata& 20.4 &\nodata& 19N, 14 & a \\
V72 & 220,h66      & 189 &\nodata& 26.4 &\nodata& 18B, 41 & a \\
V73 & 221,222,226  & 196 &\nodata& 26.4 &\nodata& 18B, 43 & a \\
V74 & 224,225,227  & 215 &\nodata& 22.1 &\nodata& 19N, 7  & s \\
V75 & 230,231      & 219 &\nodata& 23.4 &\nodata& 19N, 12 & s \\
V76 & 233u,235u    & 199 & 288   & 26.4 & 22.1  & \nodata & double  \\
V77 & 237,238      & 210 &\nodata& 17.6 &\nodata& 20O, 27 & a \\
V78 & 239,240      & 211 &\nodata& 17.6 &\nodata& 22O, 56 & a \\
V79 & 244          & 218 &\nodata& 17.6 &\nodata& 22O, 56 & s \\
V80 & 245,247,248  & 214 &\nodata& 15.8 &\nodata& 22O, 57 & a \\
V81 & 249          & 205 &\nodata& 28.1 &\nodata& \nodata & a \\
V82 & 250,251,252  & 192 &\nodata& 21.2 &\nodata& \nodata & a, broad red wing\\
V83 & h61          & 216 &\nodata& 28.1 &\nodata& 16B, 37 & a \\
V84 & nucleus      & 220 &\nodata& 30.7 &\nodata& 16B, 37 & a \\
\enddata
\end{deluxetable}

\begin{deluxetable}{cccc}
\tablewidth{0pc}
\tablecaption{ Aperture positions for search of H$\alpha$ emission
\label{HI_maxima}}
\tablehead{
\colhead{Position} & \colhead{R. A.} & \colhead{Dec}    & \colhead{Hunter} \\
\colhead{ } & \colhead{(B1950.0)} & \colhead{(B1950.0)} & \colhead{ID}
}
\startdata
1 & 12 25 04 & 43 59 11 & \nodata \\
2 & 12 25 04 & 44 14 08 & Z2      \\
3 & 12 27 05 & 44 29 08 & Z1      \\
\enddata
\tablecomments{Units for right ascension are hours, minutes and seconds.
Those for declination are degrees, arcminutes and arcseconds. The cross
identifications in column 4, are based on the work of \citet{hunter00b}. }
\end{deluxetable}

\end{document}